\documentclass[preprint2]{aastex}

\slugcomment{Not to appear in Nonlearned J., 45.}

\shorttitle{Title}
\shortauthors{Parroni et al.}

\usepackage{amsmath}
\usepackage[english]{babel} 
\usepackage[T1]{fontenc}
\usepackage[latin1]{inputenc}
\usepackage{graphicx} 
\usepackage{float} 
\usepackage{booktabs}
\usepackage{bm}
\usepackage{placeins}
\usepackage{realboxes}
\usepackage{multirow}
\usepackage[usenames,dvipsnames,svgnames,table]{xcolor}
\usepackage[titletoc,title]{appendix}
\usepackage{tabularx}
\usepackage
[
        a4paper,% other options: a3paper, a5paper, etc
        left=2cm,
        right=1cm,
        top=3cm,
        bottom=4cm,
]
{geometry}
\newcolumntype{L}[1]{>{\raggedright\let\newline\\\arraybackslash\hspace{0pt}}m{#1}}
\newcolumntype{C}[1]{>{\centering\let\newline\\\arraybackslash\hspace{0pt}}m{#1}}
\newcolumntype{R}[1]{>{\raggedleft\let\newline\\\arraybackslash\hspace{0pt}}m{#1}}

\begin{document}

\title{Next Generation Virgo Cluster Survey. XXI. The weak lensing masses of the CFHTLS and NGVS RedGOLD galaxy clusters and calibration of the optical richness}

\author{
Carolina~Parroni \altaffilmark{1,2},
Simona~Mei  \altaffilmark{1,2,3},
Thomas~Erben\altaffilmark{4},
Ludovic~Van Waerbeke\altaffilmark{5},
Anand~Raichoor\altaffilmark{6},
Jes~Ford\altaffilmark{7},
Rossella~Licitra \altaffilmark{1,2},
Massimo~Meneghetti\altaffilmark{8},
Hendrik~Hildebrandt\altaffilmark{4},
Lance~Miller\altaffilmark{9},
Patrick~C\^{o}t\'{e}\altaffilmark{10},
Giovanni~Covone\altaffilmark{11},
Jean-Charles~Cuillandre\altaffilmark{12},
Pierre-Alain~Duc\altaffilmark{13,14},
Laura~Ferrarese\altaffilmark{10},
Stephen~D.J.~Gwyn\altaffilmark{10},
%Ariane~Lan\c{c}on\altaffilmark{14},
%Roberto P.~Mu\~noz\altaffilmark{15},
Thomas H.~Puzia\altaffilmark{15}
}

\altaffiltext{1}{LERMA, Observatoire de Paris,  PSL Research University, CNRS, Sorbonne Universit\'es, UPMC Univ. Paris 06, F-75014 Paris, France}
\altaffiltext{2}{University of Paris Denis Diderot, University of Paris Sorbonne Cit\'e (PSC), F-75205 Paris Cedex 13, France}
\altaffiltext{3}{Jet Propulsion Laboratory, Cahill Center for Astronomy \& Astrophysics, California Institute of Technology, 4800 Oak Grove Drive, Pasadena, California, USA}
\altaffiltext{4}{Argelander-Institut f\"{u}r Astronomie, University of Bonn, Auf dem H\"{u}gel 71, D-53121 Bonn, Germany}
\altaffiltext{5}{Department of Physics and Astronomy, University of British Columbia, 6224 Agricultural Road, Vancouver, B.C., V6T 1Z1, Canada}
\altaffiltext{6}{Institute of Physics, Laboratory of Astrophysics, Ecole Polytechnique F\'ed\'erale de Lausanne (EPFL), Observatoire de Sauverny, 1290 Versoix, Switzerland}
\altaffiltext{7}{\textit{e}Science Institute, University of Washington, Campus Box 351570, Seattle, WA 98195, USA}
\altaffiltext{8}{INAF - Osservatorio Astronomico di Bologna, Bologna, Italy \& INFN, Sezione di Bologna, Bologna, Italy}
\altaffiltext{9}{Department of Physics, University of Oxford, Denys Wilkinson Building, Keble Road, Oxford OX1 3RH, U.K}
\altaffiltext{10}{National Research Council of Canada, Herzberg Astronomy and Astrophysics Program, 5071 West Saanich Road, Victoria, BC, V9E 2E7, Canada}
\altaffiltext{11}{INAF Osservatorio Astronomico di Capodimonte, Dipartimento di Fisica, University of Naples Federico II, INFN, Via Cinthia, I-80126 Napoli, Italy}
\altaffiltext{12}{CEA/IRFU/SAp, Laboratoire AIM Paris-Saclay, CNRS/INSU, Université Paris Diderot, Observatoire de Paris, PSL Research University, F-91191 Gif-sur-Yvette Cedex, France}
\altaffiltext{13}{AIM Paris-Saclay Service d'astrophysique, CEA-Saclay, F-91191 Gif sur Yvette, France}
\altaffiltext{14}{Universit\'{e} de Strasbourg, CNRS, Observatoire astronomique de Strasbourg, UMR 7550, F-67000 Strasbourg, France}
\altaffiltext{15}{Institute of Astrophysics, Pontificia Universidad Cat\'{o}lica de Chile, Av. Vicu\~{n}a Mackenna 4860, 7820436 Macul, Santiago, Chile}

\begin{abstract}
We measured stacked weak lensing cluster masses for a sample of 1323 galaxy clusters detected by the RedGOLD algorithm in the Canada-France-Hawaii Telescope Legacy Survey W1 and the Next Generation Virgo Cluster Survey at $0.2<z<0.5$, in the optical richness range $10<\lambda<70$. This is the most comprehensive lensing study of a $\sim 100\%$ complete and $\sim 80\%$ pure optical cluster catalog in this redshift range. We test different mass models, and our final model includes a basic halo model with a Navarro Frenk and White profile, as well as correction terms that take into account cluster miscentering, non-weak shear, the two-halo term, the contribution of the Brightest Cluster Galaxy, and an a posteriori correction for the intrinsic scatter in the mass--richness relation. With this model, we obtain a mass--richness relation of $\log{M_{\rm 200}/M_{\odot}}=(14.46\pm0.02)+(1.04\pm0.09)\log{(\lambda/40)}$ (statistical uncertainties). This result is consistent with other published lensing mass--richness relations.  We give the coefficients of the scaling relations between the lensing mass and X-ray mass proxies, $L_X$ and $T_X$, and compare them with previous results.  When compared to X-ray masses and mass proxies, our results are in agreement with most previous results and simulations, and consistent with the expected deviations from self-similarity.
\end{abstract}

\keywords{}

\section{INTRODUCTION}		
Galaxy clusters are the largest and most massive gravitationally bound systems in the universe and their number and distribution permit us to probe the predictions of cosmological models. They are the densest environments where we can study galaxy formation and evolution, and their interaction with the intra-cluster medium \citep{voit2005}. For both these goals, an accurate estimate of the cluster mass is essential. 
				
The cluster mass cannot be measured directly, but is inferred using several mass proxies. Galaxy clusters emit radiation at different wavelengths and their mass can be estimated using different tracers. Different mass proxies usually lead to mass estimations that are affected by different systematics.
 
From X-ray observations of the cluster gas, we can derive the gas temperature, which is related to its total mass \citep{sarazin1988}, under the assumption of hydrostatic equilibrium. X-ray mass measurements are less subjected to projection and triaxiality effects, but the mass proxies are not reliable in systems undergoing mergers or in the central regions of clusters with strong AGN feedback \citep{allen2011}.

The intracluster medium (ICM) can also be detected in the millimeter by the thermal Sunyaev--Zel'dovich effect \citep[S-Z effect;][]{sz1972} and the S-Z flux is related to the total cluster mass. Unlike optical and X-ray surface brightness, the integrated S-Z flux is independent of distance, allowing for almost constant mass limit measurements at high redshifts. For the same reason, though, the method is also subjected to projection effects due to the overlap of all the groups and clusters along the line of sight \citep{voit2005}.

In the optical and infrared bandpasses, we observe the starlight from cluster galaxies. If a cluster is in dynamical equilibrium, the velocity distribution of its galaxies is expected to be Gaussian and the velocity dispersion can be directly linked to its mass through the virial theorem. An advantage of this method is that, unlike X-ray and S-Z mass measurements, it is not affected by forms of non-thermal pressure such as magnetic fields, turbulence, and cosmic ray pressure. On the downside, it is sensitive to triaxiality and projection effects, the precision of the measurements is limited by the finite number of galaxies, and the assumption of dynamical and virial equilibrium is not always correct \citep{allen2011}.

The total optical or infrared luminosity of a cluster is another indicator of its mass, given that light traces mass. \citet{abell1958} defined a \textit{richness} class to categorize clusters based on the number of member galaxies brighter than a given magnitude limit. The luminosity distribution function of cluster galaxies is also well described by the \citet{schechter1976} profile, and the observation of the high-luminosity tip of this distribution allows us to better constrain cluster masses. \citet{postman1996}, for example, defined the richness parameter as the number of cluster galaxies brighter than the characteristic luminosity of the \citet{schechter1976} profile, $L_{\rm *}$. Different definitions are possible and intrinsically related to the technique used to optically detect galaxy clusters. 

\citet{rykoff2014} built an optical cluster finder based on the red-sequence finding technique, redMaPPer and applied it to the Sloan Digital Sky Survey \citep[SDSS;][]{york2000}. Their richness is computed using optimal filtering as a sum of probabilities and depends on three filters based on colors, positions, and luminosity \citep{rozo2009, rykoff2012, rykoff2014, rykoff2016, rozo2014}.

In \citet{licitra2016a,licitra2016b}, we introduced a simplified definition of cluster richness based on the redMaPPer richness measurement, within our detection and cluster selection algorithm RedGOLD. RedGOLD is based on a revised red-sequence technique.

RedGOLD richness quantifies the number of red, passive, early-type galaxies (ETGs) brighter than $0.2L_{\rm *}$, inside a scale radius, subtracting the scaled background. When compared to X-ray mass proxies, the RedGOLD richness leads to scatters in the X-ray temperature-richness relation similar to those obtained with redMaPPer \citep{rozo2014}, which is very promising because RedGOLD was applied to a lower richness threshold (i.e. lower cluster mass).

The total cluster mass can also be derived by its strong and weak gravitational lensing of background sources. In the weak lensing regime, the gravitational potential of clusters of galaxies produces small distortions in the observed shape of the background field galaxies, creating the so-called shear field, which is proportional to the cluster mass.  

Because the shear is small relative to the intrinsic ellipticity of the galaxies (due to their random shape and orientation), a statistical approach is required to measure it and the signal is averaged over a large number of background sources to increase the signal-to-noise ratio \citep[S/N; ][]{schneider2006}. Gravitational lensing does not require any assumptions about the dynamical state of the cluster and it is sensitive to the projected mass along the line of sight, making it a more reliable tool to determine total cluster masses \citep{meneghetti2010, allen2011, rasia2012}.

In the future, as shown in \citet{ascaso2016}, optical and near-infrared (NIR) cluster surveys, such as Euclid\footnote{http://euclid- ec.org} \citep{laureijs2011}, Large Synoptic Survey Telescope (LSST)\footnote{http://www.lsst.org}  and J-PAS \citep{benitez2014}, will reach deeper than X-ray and S-Z surveys, such as e-Rosita \citep{merloni2012}, SPTpol \citep{carlstrom2011} and ACTpol \citep{marriage2011}. It is thus important to understand the reliability of optical and NIR mass proxies because they will be the only mass proxy available for these new detections.
		
Several works in the literature have proven that the optical richness shows a good correlation with the cluster total masses derived from weak lensing \citep{johnston2007,covone2014,ford2015,vanuitert2015, melchior2016, simet2016}. From these works, the typical uncertainty found in the cluster mass at a given richness is $\sim10-25\%$ including statistical and systematic errors, in the mass range $6 \times 10^{12}M_{\rm \odot} \lesssim M \lesssim 10^{15}M_{\rm \odot}$ and in the redshift range $0.1\lesssim z \lesssim 0.9$.

The aim of this work is to calibrate and evaluate the precision of the RedGOLD richness as a mass proxy and to compare it to stacked weak-lensing masses. We then compare our lensing-calibrated masses to X-ray mass proxies. Our approach mainly follows the one adopted by \citet{johnston2007} and \citet{ford2015}, and we compare our results to \citet{simet2016}, \citet{farahi2016} and \citet{melchior2016}.

The paper is organized as follows: in Section 2, we describe the shear data set and the photometric redshifts catalog; in Section 3, we briefly present the RedGOLD detection algorithm and the cluster catalogs; in Section 4, we describe the weak-lensing equations and our method; in Section 5, we present our results; in Section 6, we discuss our findings in comparison with other recent works; in Section 7, we present our conclusions.
		    
Throughout this work, we assume a standard $\Lambda CDM$ model, with $\Omega_{\rm {m}}=0.3$, $\Omega_{\rm \Lambda}=0.7$ and $\rm{H_{\rm 0}}=70~km~s^{-1}~Mpc^{-1}$.

Magnitudes are given in the AB system \citep{oke1983,sirianni2005}. 

\section{DATA}

For our analysis, we use our own data reduction \citep{raichoor2014} of the Canada-France-Hawaii Telescope
Legacy Survey \citep[CFHT-LS;][]{gwyn2012} Wide~1 (W1) field and of the Next Generation Virgo Cluster Survey \citep[NGVS;][]{ferrarese2012}. We describe these two data sets below.

	\subsection{CFHTLenS and NGVSLenS}\label{shear_data}
		
	The CFHT-LS is a multi-color optical survey conducted between 2003 and 2008 using the CFHT optical multi-chip MegaPrime instrument \citep[MegaCam\footnote{http://www.cfht.hawaii.edu/Instruments/Imaging/ Megacam/};][] {boulade2003}. The survey consists of 171 pointing covering $\sim154~deg^{2}$ in four wide fields ranging from 25 to 72 $\rm deg^{2}$, with complete color coverage in the five filters $u^{*}g^{\prime}r^{\prime}i^{\prime}z^{\prime}$. All the pointings selected for this analysis were obtained under optimal seeing conditions with a seeing $<0.8^{\arcsec}$ in the primary lensing band $i^{\prime}$ \citep{erben2013}. The $5\sigma$ point source limiting magnitudes in a $2.0^{\arcsec}$ aperture in the five $u^{*}g^{\prime}r^{\prime}i^{\prime}z^{\prime}$ filters are $\sim25.2$, $\sim25.6$, $\sim24.9$, $\sim24.5$, $\sim23.5$~mag, respectively \citep{erben2013}. 
		
	The NGVS \citep{ferrarese2012} is a  multi-color optical imaging survey of the Virgo Cluster, also obtained with the CFHT MegaCam instrument. This survey covers $104~deg^{2}$ with 117 pointings  in the four filters $u^{*}g^{\prime}i^{\prime}z^{\prime}$. Thirty-four of these pointings are also covered in the $r^{\prime}$ band. As for the CFHT-LS, the optimal seeing conditions were reserved to the $i^{\prime}$-band, which covers the entire survey with a seeing $<0.6^{\arcsec}$. The $5\sigma$ point source limiting magnitudes in a $2.0^{\arcsec}$ aperture in the five $u^{*}g^{\prime}r^{\prime}i^{\prime}z^{\prime}$ filters are $\sim25.6$, $\sim25.7$, $\sim24.7$, $\sim24.4$, $\sim23.6$~mag, respectively \citep{raichoor2014}.
		
	 Both our CFHTLenS and NGVSLenS photometry and photometric redshift catalogs were derived using the dedicated data processing described in \citet{raichoor2014}. The preprocessed \textit{Elixir}\footnote{http://www.cfht.hawaii.edu/Instruments/Elixir/} data, available at the Canadian Astronomical Data Centre (CADC)\footnote{ http://www4.cadc-ccda.hia-iha.nrc-cnrc.gc.ca/cadc/}) were processed with an improved version of the THELI pipeline \citep{erben2005, erben2009, erben2013, raichoor2014} to obtain co-added science images accompanied by weights, flag maps, sum frames, image masks, and sky-subtracted individual chips that are at the base of the shear and photometric analysis. We refer the reader to \citet{erben2013} and \citet{heymans2012}  for a detailed description of the different THELI processing steps and a full systematic error analysis. \citet{raichoor2014} modified the standard pipeline performing the zero-point calibration using the SDSS data, taking advantage of its internal photometric stability. The SDSS covers the entire NGVS field and 62 out of 72 pointings of the CFHT-LS W1 field ($\sim 60~deg^{2}$). \citet{raichoor2014} constructed the photometric catalogs as described in \citet{hildebrandt2012}, adopting a {\it global} PSF homogenization to measure unbiased colors. Multicolor catalogs were obtained from PSF-homogenized images using SExtractor \citep{bertin1996} in dual-image mode, with the un-convolved $i^{\prime}$-band single-exposure as the detection image.
	 
	 We restrict our analysis to the entire NGVS and the $\sim60~deg^2$ of the W1 field that were reprocessed by \citet{raichoor2014}, to have an homogeneously processed photometric catalog on a total of $\sim164~deg^2$.

	For the shear analysis, as described in \citet{miller2013}, shape measurements were obtained applying the Bayesian \textit{lens}fit algorithm to single-exposure $i^{\prime}$-band images with accurate PSF modeling, fitting PSF-convolved disc plus bulge galaxy models. The ellipticity of each galaxy is estimated from the mean likelihood of the model posterior probability, marginalized over model nuisance parameters of galaxy position, size, brightness, and bulge fraction. The code assigns to each galaxy an inverse variance weight $w_{\rm {lens}} \propto (\sigma_{\rm{e}}^{2}+\sigma_{\rm {pop}}^{2})^{-1}$, where $\sigma_{\rm {e}}^{2}$ is the variance of the ellipticity likelihood surface and $\sigma_{\rm {pop}}^{2}$ is the variance of the ellipticity distribution of the galaxy population. Calibration corrections consist of a multiplicative bias $m$, calculated using simulated images, and an additive bias $c$, estimated empirically from the data. As discussed in \citet{miller2013}, the former increases as the size and the S/N of a galaxy detection decrease, while the latter increases as the S/N of a galaxy detection increases and the size decreases.

	\subsection{Photometric Redshifts}\label{sec:photoz}
			
	The photometric redshift catalogs of the $\sim 60~deg^{2}$ of the CFHTLenS covered by the SDSS and of the entire NGVSLenS were obtained using the Bayesian software packages \textit{LePhare} \citep{arnouts1999, arnouts2002, ilbert2006} and BPZ \citep{benitez2000, benitez2004, coe2006}, as described in \citet{raichoor2014}. We used the re-calibrated SED template set of \citet{capak2004}. 
	
	Both {\it LePhare} and BPZ are designed for high-redshift studies, giving biased or low-quality photo-z's estimations for objects with $i^{\prime}<20$~mag, which represent a non-negligible fraction of both samples. In order to improve the performance at low redshift, \citet{hildebrandt2012} used an \textit{ad hoc} modified prior for the CFHTLenS data. \citet{raichoor2014} adopted a more systematic solution for our reprocessed CFHTLenS W1 field and for the NGVSLenS, building a new prior calibrated on observed data, using the SDSS Galaxy Main Sample spectroscopic survey \citep{york2000, strauss2002, ahn2014} to include bright sources.

	To analyze the accuracy of the photometric redshift estimates, \citet{raichoor2014} used several spectroscopic surveys covering the CFHTLenS and NGVSLenS: the SDSS Galaxy Main Sample, two spectroscopic programs at the Multiple Mirror Telescope (MMT; Peng, E. W. et al. 2016, in preparation) and at the Anglo-Australian Telescope \citep[AAT;][2016, in preparation]{zhang2015}, the Virgo Dwarf Globular Cluster Survey (Guhathakurta, P. et al. 2016, in preparation), the DEEP2 Galaxy Redshift Survey over the Extended Groth Strip \citep[DEEP2/EGS;][]{davis2003, newman2013}, the VIMOS Public Extragalactic Redshift Survey \citep[VIPERS;][]{guzzo2014}, and the F02 and F22 fields of the VIMOS VLT Deep Survey \citep[VVDS;][]{lefevre2005, lefevre2013}. 
		
	As shown in \citet{raichoor2014}, when using all five filters, for $0.2<z_{\rm phot}<1$ and $i^{\prime}<23$~mag, we found a bias $\Delta z=\frac{z_{\rm phot}-z_{\rm spec}}{1+z_{\rm spec}}<0.02$ with scatter values in the range $0.02<\sigma<0.05$ and $<5\%$ of outliers. When using four bands, the quality of the measurements slightly decreases, due to the lack of the $r^{\prime}$-band to sample the 4000~{\AA} break. In the range $0.3<z_{\rm phot}<0.8$ and $i^{\prime}>21$~mag, we obtained $-0.05<bias<0.02$, a scatter $\sigma\sim0.06$ and an outlier rate of $10-15\%$. Our photometric redshifts are not reliable for $z<0.2$ \citep{raichoor2014} and we excluded these low redshifts from our cluster detection in \citet{licitra2016a,licitra2016b} and our weak lensing analysis.
	
	In this analysis, we use the photometric redshifts derived with BPZ, corresponding to $z_{\rm best}$, the peak of the redshift posterior distribution (hereafter, $z_{\rm phot}$).
	
\section{CLUSTER CATALOGS}

\subsection{The RedGOLD Optical Cluster Catalogs}\label{redgold}
	\subsubsection{The RedGOLD Algorithm}
		
		The RedGOLD algorithm \citep{licitra2016a, licitra2016b} is based on a modified red-sequence search algorithm. Because the inner regions of galaxy clusters host a large population of passive and bright ETGs, RedGOLD searches for passive ETG overdensities.
	 To avoid the selection of dusty red star-forming galaxies, the algorithm selects galaxies on the red sequence both in the  rest-frame $(U - B)$ and $(B - V)$, using red sequence rest-frame zero point, slope, and scatter from \citet{mei2009}, as well as an ETG spectral classification from \textit{LePhare}. In order to select an overdensity detection as a cluster candidate, the algorithm also imposes that the ETG radial distribution follows an NFW \citep{NFW1996} surface density profile.

		RedGOLD centers the cluster detection on the ETG with the highest number of red companions, weighted by luminosity. This is motivated by the fact that the brightest cluster members lying near the X-ray centroid are better tracers of the cluster centers compared to using only the BCG \citep{george2012}. The redshift of the cluster is the median photometric redshift of the passive ETGs.
		
		Each detection is characterized by two parameters--the significance $\sigma_{\rm det}$ and the richness $\lambda$--which quantifies the number of bright red ETGs inside the cluster, using an iterative algorithm. 	
		
		The entire galaxy sample is divided into overlapping photometric redshift slices. Each slice is then divided in overlapping circular cells, with a fixed comoving radius of $500~kpc$. The algorithm counts $N_{\rm gal}$, the number of red ETGs inside each cell, brighter than $0.2L_{\rm *}$, building the galaxy count distribution in each redshift slice. The background contribution is defined as $N_{\rm bkg}$, the mode of this distribution, with standard deviation $\sigma_{\rm bkg}$. The detection significance is then defined as $\sigma_{\rm det}=(N_{\rm gal}-N_{\rm bkg})/\sigma_{\rm bkg}$. Overdensities larger than $N_{\rm bkg}+\sigma_{\rm det}\times \sigma_{\rm bkg}$ are selected as preliminary detections. The uncertainties on the cluster photometric redshift range between 0.001 and 0.005, with an average of $0.003\pm0.002$. In this paper, we assume that these uncertainties are negligible for our analysis \citep[see also][]{simet2016}.
		
		The algorithm then estimates the richness $\lambda$, counting $N_{\rm gal}$ inside a scale radius, initially set to $1~Mpc$. The radius is iteratively scaled with richness as in \citet{rykoff2014}, until the difference in richness between two successive iterations is less than $N_{\rm bkg}$.
		
		RedGOLD is optimized to produce cluster catalogs with high completeness and purity. In \citet{licitra2016a, licitra2016b}, the \textit{completeness} is defined as the ratio between detected structures corresponding to true clusters and the total number of true clusters, and the \textit{purity} is defined as the number of detections that correspond to real structures to the total number of detected objects.
		
		Following the definition of a \textit{true cluster} in the literature  \citep[e.g, ][]{finoguenov2003, lin2004, evrard2008, finoguenov2009, mcgee2009, mead2010, george2011, chiangoverzier2013, gillis2013, shankar2013}, we define a \textit{true cluster}  as a dark matter halo more massive than $10^{14}\ \rm M_{\odot} $. In fact, numerical simulations show that 90$\%$  of the dark matter halos more massive than $10^{14}\ M_{\odot}$ are a very regular virialized cluster population up to redshift $z\sim1.5$  \citep[e.g.,][]{evrard2008, chiangoverzier2013}. In order to validate the performance of our algorithm to find clusters with a total mass larger than $10^{14}\ \rm M_{\odot} $ and measure our obtained sample completeness and purity, we have applied RedGOLD to both galaxy mock catalogs and observations of  X-ray detected clusters \citep{licitra2016a}. For  details on the method and the performance of the algorithm when applied to simulations and observations, we refer the reader to \citet{licitra2016a}.
	
	\subsubsection{The RedGOLD CFHT-LS W1 and NGVS Cluster Catalogs}\label{redgold_cat}

		We use the CFHT-LS W1 and NGVS cluster catalogs from \citet{licitra2016a} and \citet{licitra2016b}, respectively. For both surveys, when using five bandpasses, in the published catalogs, we selected clusters more massive than $\approx 10^{14}M_{\rm \odot}$, the mass limit for which $\sim 90\%$ of dark matter halos at $z_{\rm phot}<1.5$ are virialized \citep{evrard2008}. In \citet{licitra2016a,licitra2016b}, we calibrated the $\sigma_{\rm det}$ and $\lambda$ parameters to maximize the completeness and purity of the catalog of these type of objects.

		\citet{licitra2016a} demonstrated that, when we considered only detections with $\sigma_{\rm det}\ge4$ and $\lambda\ge10$ at $z_{\rm phot}\le0.6$, and $\sigma_{\rm det}\ge4.5$ and $\lambda\ge10$ at $z_{\rm phot}\lesssim1$, we obtain catalogs with a completeness of $\sim100\%$ and $\sim70\%$, respectively, and a purity of $\sim80\%$ \citep[see Figure~7 and 8 from][]{licitra2016a}.
					
		In both the CFHT-LS W1 and the NGVS, we masked areas around bright stars and nearby galaxies. We found that in only $\sim2\%$ of the cluster candidates (low richness structures at high redshift) are more than $10\%$ of their bright potential members masked \citep{licitra2016a}. Therefore, our richness estimates are not significantly affected by masking.
		
		For the NGVS, as explained above, the five-band coverage was limited to only the $\sim30\%$ of the survey. The lack of the $r^{\prime}$-band in the remaining pointings, causes higher uncertainties on the determination of photometric redshifts for sources at $0.3<z_{\rm phot}<0.8$ but the global accuracy on the photometric redshifts remains high even for this sample, as shown in \citet{raichoor2014}. Because there are some fields in which the quality of the $r^{\prime}$-band is lower because of the lower depth and the lack of coverage of the intra-CCD regions, this adds to the difficulty of detecting the less-massive structures at intermediate and high redshifts, as well as the determination of the clusters center and richness.

		To quantify this effect in the richness estimation, \citet{licitra2016b} compared the values recovered with a full band coverage $\lambda_{\rm r}$ to the ones obtained without the $r^{\prime}$-band $\lambda_{\rm wr}$, and measured  $\Delta \lambda/\lambda_{\rm r} \equiv (\lambda_{\rm r}-\lambda_{\rm wr})/\lambda_{\rm r}$, in different redshift bins. Median values of $\Delta \lambda/\lambda_{\rm r}$ and their standard deviations are listed in Table~2 of \citet{licitra2016b}. At $z_{\rm phot}<0.5$ and $z_{\rm phot}>0.8$, the two estimates are in good agreement, with $\Delta \lambda/\lambda_{\rm r}<10\%$. This is due to the fact that  the $(g-z)$ and $(i-z)$ colors straddle the 4000~{\AA} break at $z_{\rm phot}<0.5$ and $z_{\rm phot}>0.8$, respectively. At $0.5<z_{\rm phot}<0.6$, $\lambda_{\rm wr}$ is systematically underestimated of $\sim40\%$ on average and, at $0.6<z_{\rm phot}<0.8$, it is systematically overestimated of $\sim20\%$ on average. The first systematic is due to the use of the $(g-z)$ color, which changes less steeply with redshift and has larger photometric errors, compared with $(r-i)$ and $(i-z)$ colors. The latter is caused by the use of the $(i-z)$ color only, without the additional cut in the $(r-z)$ or $(r-i)$ colors that allows us to reduce the contamination of dusty red galaxies on the red sequence \citep{licitra2016b}.
		
		To take this into account, we correct the $\lambda_{\rm wr}$ estimations using the average shifts given in Table~2 of \citet{licitra2016b}. As we will discuss later, since for this analysis we are only selecting clusters at $z_{\rm phot}<0.5$ (see below), so using four bands preserves the same level of completeness and purity as using the five-bands catalog. 
		
		\begin{figure*}[!htbp]
                \centering
                \includegraphics[scale=.50]{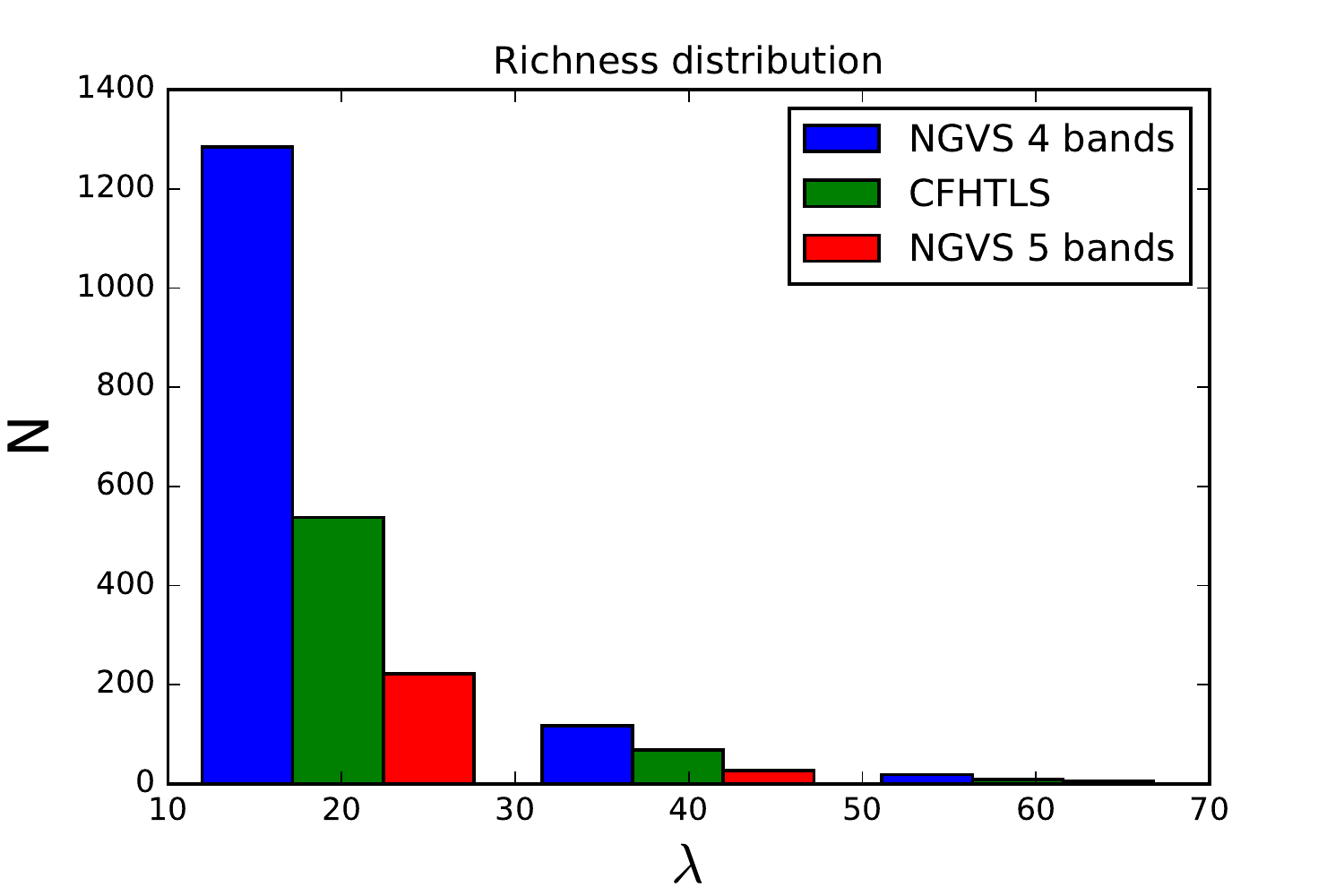}
                \includegraphics[scale=.50]{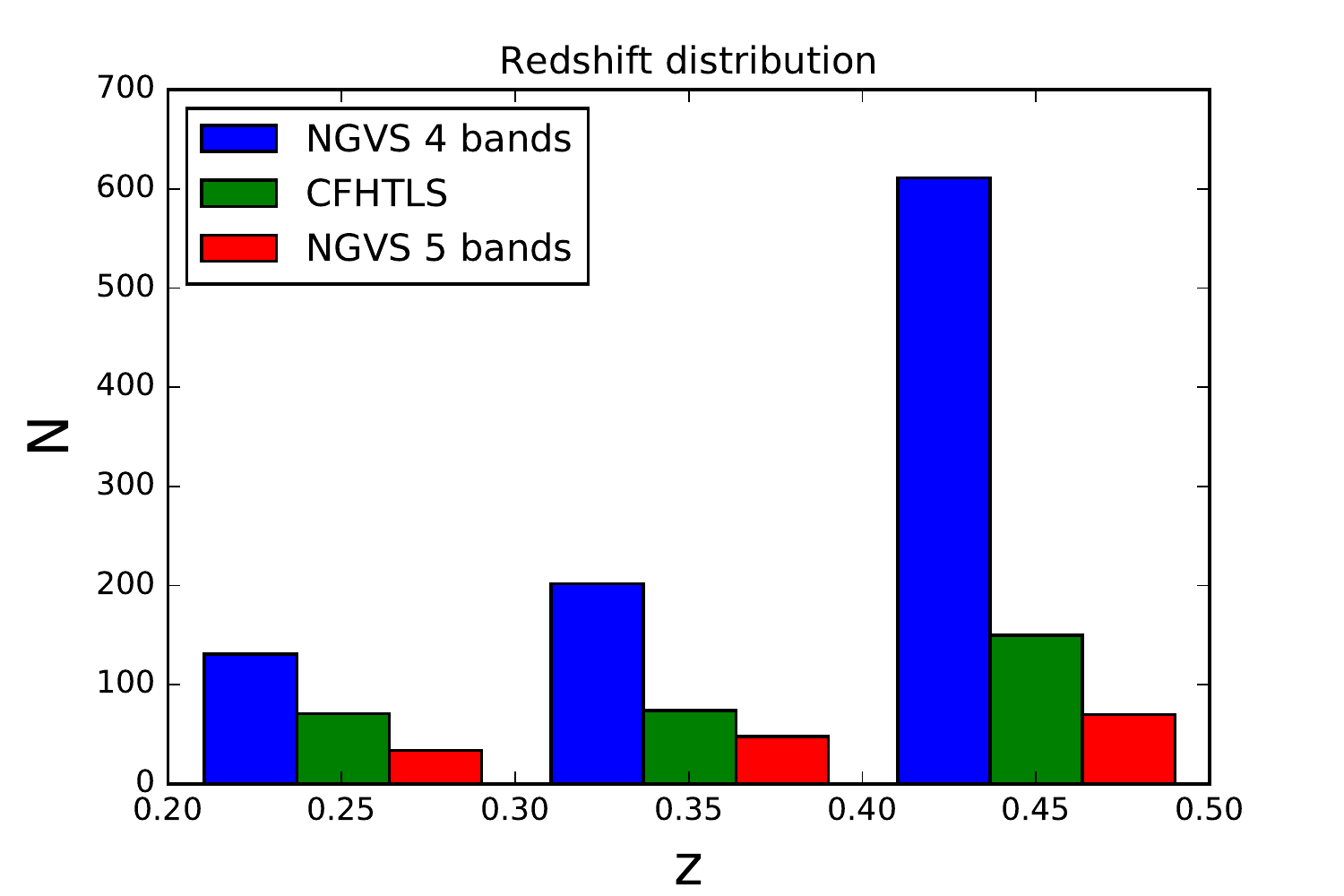}
                \caption{\scriptsize The richness and redshift distributions of the  RedGOLD CFHT-LS W1, NGVS5, and NGVS4 1323 clusters from published catalogs, selected for our weak lensing analysis (see text for the description of the catalogs).  The richness is plotted in bins of $\Delta \lambda=20$, and the redshift in bins of $\Delta z=0.1$. In each bin, the bars corresponding to the three different samples are plotted next to each other.}
                \label{fig:isto}
                \end{figure*}
		
		For these reasons, we built two separate catalogs for the NGVS: the first for the $\sim20~deg^{2}$ covered by the $r^{\prime}$-band and the second for the entire NGVS using only four bandpasses. In this last catalog, we corrected for the average shift in $\lambda$ when applying our thresholds \citep{licitra2016b}. Hereafter, we define the NGVS catalog obtained on the area covered by the five bandpasses as NGVS5 and the catalog obtained with four bandpasses as NGVS4.
		
		The CFHT-LS W1 published catalog includes 652 cluster candidate detections in an area of  $\sim60~deg^{2}$. The NGVS published catalogs include 279 and 1505 detections, in the $\sim20~deg^{2}$ with the five band coverage and in the rest of the survey, respectively. 
		
		We select cluster subsamples from these catalogs for our weak lensing analysis. Knowing that the peak in the lensing efficiency is found at  $z_{\rm phot}\sim0.4$ for source galaxies at $z_{\rm phot}\sim1$ \citep{hamana2004} and that shear measurements from ground-based telescopes are reliable for clusters with redshifts $0.2<z_{\rm phot}<0.5$ \citep{kasliwal2008}, we select detections only in this redshift range $0.2<z_{\rm phot}<0.5$, where the lower limit is due to the fact that our photometric redshifts are not reliable for $z_{\rm phot}<0.2$, as noted in Section \ref{sec:photoz} and presented in \citet{raichoor2014} and \citet{licitra2016a}. We also discard clusters with richness $\lambda<10$ and $\lambda>70$. In fact, as shown in  \citet{licitra2016a} at richness $\lambda<10$, our purity decreases for a given significance threshold. For our significance threshold of $\sigma_{\rm det} >4$, $\lambda<10$ implies a contamination of false detections larger than $\sim 20\%$. For $\lambda>70$, we have very few detections and there are not enough clusters to obtain an average profile from a statistically significant sample. 
		
		Our final selection for the weak lensing analysis includes 1323 clusters. Their richness and redshift distributions are shown in Figure \ref{fig:isto}. Hereafter, we will define the catalogs to which we applied the thresholds in significance, richness, and redshift for the weak lensing analysis as {\it selected} catalogs. The published \citet{licitra2016a, licitra2016b} catalogs, to which we applied the thresholds in significance and richness, will be referred to as Licitra's {\it published} catalogs. The \citet{licitra2016a, licitra2016b} catalogs, without any threshold, will be called {\it complete} catalogs. 
			
		\subsection{The X-Ray Cluster Catalogs}\label{xray_cat}
		
	 		\citet{gozaliasl2014} analyzed the {\it XMM-Newton} observations in the $\sim 3~deg^{2}$ overlapping the CFHT-LS W1 field, as a part of the XMM-LSS survey \citep{pierre2007} \footnote{https://heasarc.gsfc.nasa.gov/W3Browse/all/cfhtlsgxmm.html}. They presented a catalog of 129 X-ray groups, in a redshift range $0.04<z_{\rm phot}<1.23$, characterized by a rest frame $0.1-2.4~keV$ band luminosity range $10^{41}-10^{44}~ergs~s^{-1}$. They removed the contribution of AGN point sources from their flux estimates and applied a correction of $\sim 10\%$ for the removal of cool core flux based on the high-resolution {\it Chandra} data on COSMOS as shown in \citet{leauthaud2010}. They used a two-color red-sequence finder to identify group members and calculate the mean group photometric redshift. They inferred cluster's $M_{\rm 200}$ masses using the $L_{\rm X}-M$ relation of \citet{leauthaud2010}, with a systematic uncertainty of $\sim 20\%$.
		
			\citet{mehrtens2012} presented the first data release of the XMM Cluster Survey, a serendipitous search for galaxy clusters in the {\it XMM-Newton} Science Archive data \footnote{https://heasarc.gsfc.nasa.gov/W3Browse/all/xcs.html}. The catalog consists of 503 optically confirmed clusters, in a redshift range $0.06<z_{\rm phot}<1.46$. Four hundred and two of these clusters have measured X-ray temperatures in the range $0.4<T_{\rm X}<14.7~keV$. They derived photometric redshifts with the red-sequence technique, using one color. They used a spherical $\beta$-profile model \citep{cavaliere1976} to fit the surface brightness profile and derive the bolometric (0.05 - 100 keV band) luminosity in units of $10^{44}~erg~s^{-1}$ within the radius $R_{\rm 200}$ and $R_{\rm 500}$.
		
			In Section \ref{xray}, we will use these catalogs to compare our lensing masses with X-ray masses and calculate the scaling relations between lensing masses and X-ray temperature and luminosity. We analyze the two catalogs separately because the different treatment of the emission from the central regions of the clusters leads to different mass estimates. In Section \ref{xray2} we will discuss these results. 
		
		\section{WEAK LENSING ANALYSIS}
		
		In this section, we describe our weak lensing analysis.
		Our aim is to infer cluster masses by reconstructing the tangential shear radial profile $\gamma_{\rm t}(R)$, averaging in concentric annuli around the halo center, and fitting it to a known density profile. 
		Here, $\gamma_{\rm t}(R)$ accounts for the distortion, due to the gravitational potential of the lens, of the shape of the background sources in the tangential direction with respect to the center of the lens. It is defined as:
		\begin{equation}
			\gamma_{\rm t}=-Re\left[\gamma e^{-2i\phi}\right]
		\end{equation}
		with $\gamma=\epsilon_{\rm 1}+i\epsilon_{\rm 2}=|\gamma|e^{2i\phi}$, where $\epsilon_{\rm 1}$ and $\epsilon_{\rm 2}$ are the ellipticity components of the galaxy and $\phi$ is the position angle of the galaxy respect to the center of the lens \citep{schneider2006}.
		
		As described in \citet{wright2000}, the tangential shear profile $\gamma_{\rm t}(R)$ is related to the surface density contrast by:
			\begin{equation}
				\Delta\Sigma(R)=\left<\gamma_{\rm t}(R)\right>{\Sigma_{\rm c}}
			\end{equation}
			where $R$ is the projected radius with respect to the center of the lens and:
			\begin{equation}\label{sigmac}
				\Sigma_{\rm c}=\frac{c^{2}}{4\pi G}\frac{D_{\rm s}}{D_{\rm l}D_{\rm ls}}
			\end{equation}
			is the critical surface density. Here, $c$ is the speed of light and $D_{\rm s}$, $D_{\rm l}$, and $D_{\rm ls}$ are the angular diameter distances from the observer to the source, from the observer to the lens, and from the lens to the source, respectively.
			
		To infer cluster masses, we fit the measured $\Delta \Sigma(R)$ profile, obtained as described in Section \ref{measured_prof}, to the theoretical models introduced in Section \ref{model_prof}.

		\subsection{Cluster Profile Measurement}\label{measured_prof}
		
			To measure cluster masses, we need to fit the cluster radial profiles. This is possible individually only for the most massive clusters in our sample ($M_{\rm 200}> 4\times 10^{14}M_{\rm \odot}$ for a signal-to-noise ratio $S/N > 3$; they represent the $\sim 2\%$ of the sample), while the noise dominates for the others. In order to increase the S/N and measure average radial profiles for all the other detections, we stack galaxy clusters in five richness bins, from $\lambda =10$ to $\lambda =70$, in steps of 10 (20 for the last bin) in richness.
		
			We select the background galaxy sample using the following criteria:
			
				\begin{multline}
					z_{\rm phot, s}>z_{\rm phot, l}+3\times \sigma_{\rm z_{phot}}\left(i^\prime-mag_{\rm s} \right)\\\times \left(1+z_{\rm phot, s} \right)	
				\end{multline}
			
			where $z_{\rm phot, s}$ is the source redshift, $z_{\rm phot, l}$ is the lens redshift, and $\sigma_{\rm z_{phot}}\left(i^\prime-mag_{\rm s}\right)$ is the error on the photometric redshift as a function of the source $i^\prime$-band magnitude. This function was obtained by interpolating the values in Figure 9 of \citet{raichoor2014}, up to $i^\prime \sim 24.7~mag$. We tested different cuts in magnitude ($i^\prime \sim 24.7, ~24, ~23.5, ~23~mag$), and found consistent results in all cases. We can conclude that the inclusion of faint sources in the background sample does not introduce a bias in the total cluster mass estimation.

			Following \citet{ford2015}, we then sort the background galaxies in 10 logarithmic radial bins from $0.09~Mpc$ from the center of the lens to $5~Mpc$.  In fact, at radii closer than 0.09~$Mpc$, galaxy counts are dominated by cluster galaxies, and at larger radii, the scatter in the mass estimate can be $\geq 20\%$ because of the contribution of large-scale structure \citep{becker2011, oguri2011}.
					
			In each radial bin, we perform a weighted average of the lensing signal as follows:
			
				\begin{equation}\label{eq:mean1}
					\Delta\Sigma(R)=\frac{{\sum_{i=0}^{l}}{\sum^s_{j=0}}w_{ij}\Sigma_{{\rm c},ij}\gamma_{{\rm t},ij}}{{\sum^l_{i=0}}{\sum^s_{j=0}} w_{ij}}
				\end{equation}
			where we sum over every \textit{lens-source} pair (i.e. {\it i--j} indices up to the $l$ number of lenses and $s$ number of sources). The weights $w_{ij}=\Sigma^{-2}_{{\rm c},ij}w_{\rm lens}$ \citep{mandelbaum2005} quantify the quality of the shape measurements through the \textit{lens}fit weights $w_{\rm lens}$ (defined in Section \ref{shear_data}) and down-weight source galaxies that are close in redshift to the lens through $\Sigma^{-2}_{{\rm c},ij}$, which is evaluated for every \textit{lens-source} pair using $z_{\rm phot}$ to calculate the angular diameter distances that appear in Equation \ref{sigmac}.

	 			We then need to correct the measured signal, applying the calibration corrections introduced in Section \ref{shear_data}. As shown in \citet{heymans2012}, the ellipticity estimated by {\it lens}fit can be related to the true ellipticity (i.e. the sum of the shear and of the galaxy intrinsic ellipticity) as $\epsilon_{\rm lens}=(1+m)[\gamma + \epsilon_{\rm int}] +c$, where $m$ and $c$ are the multiplicative and additive biases. While the latter can be simply added on single ellipticity measurements, the first needs to be applied as a weighted ensemble average correction:
			
				\begin{equation}
					1+K(R) \equiv \frac{\sum_{i=0}^{l}\sum_{j=0}^{s}w_{ij}[1+m_{ij}]}{\sum_{i=0}^{l}\sum_{j=0}^{s}w_{ij}}
				\end{equation}
		
			This is done to avoid possible instabilities in case the term $(1+m)$ tends to zero. In this way, we also remove any correlation between the calibration correction and the intrinsic ellipticity \citep{miller2013}. The calibrated signal is written as:
			
				\begin{equation}\label{eq:mean}
					\Delta\Sigma_{\rm cal}(R)=\frac{\Delta\Sigma(R)}{1+K(R)}
				\end{equation}
				
				To estimate the errors on $\Delta\Sigma(R)$, we create a set of 100 bootstrap realizations for each richness bin, selecting the same number of clusters for each stack but taking them with replacements. We apply Equation \ref{eq:mean1} to obtain $\Delta\Sigma(R)$ for each bootstrap sample.
				
				Following  \citet{ford2015}, we then calculate the covariance matrix:
				\begin{multline}
					C(R_{i},R_{j})=\\ \left[\frac{N}{N-1}\right]^{2}\frac{1}{N} \sum_{k=1}^{N} \left[ \Delta\Sigma_{k}(R_{i})-\overline{\Delta\Sigma}(R_{i})\right]\\ \times \left[ \Delta\Sigma_{k}(R_{j})-\overline{\Delta\Sigma}(R_{j})\right]
					\label{eq:cov}
				\end{multline}
				where $R_{i}$ and $R_{j}$ are the radial bins, $N$ is the number of bootstrap samples, and $\overline{\Delta\Sigma}(R_{i})$ is the average over all bootstrap realizations. 
				
				 For each radial bin, we weight the shear using the \textit{lens}fit weights as shown in Equation \ref{eq:mean1}, so these error bars also include the error on the shape measurements of the source galaxies. We calculate the covariance matrix to take into account the correlation between radial bins and the contribution to the stacked signal of clusters with different masses inside the same richness bin.

		\subsection{Cluster Profile Model}\label{model_prof}

		In order to fit the tangential shear profiles, we use a basic analytic model for the cluster profile, to which we progressively add additional terms to obtain our fiducial model, which we will call {\it Final model}. This procedure permits us to quantify how adding additional terms changes the final cluster profile model.
	
	Our basic analytic model is the following (hereafter {\it Basic Model}): 
			
			\begin{multline}\label{deltasigmatot}
				\Delta\Sigma(R)=p_{\rm cc}[ \Delta\Sigma_{\rm NFW}(R)+ \Delta\Sigma_{\rm nw}(R)]+ \\ (1-p_{\rm cc})\Delta\Sigma_{\rm sm}(R)+\Delta\Sigma_{\rm 2halo}(R)
			\end{multline}	
				
		 	Here, $\Delta\Sigma_{\rm NFW}$ is the surface density contrast calculated from an NFW density profile, assumed as the halo profile; $\Delta\Sigma_{\rm nw}$, $\Delta\Sigma_{\rm sm}$ and $\Delta\Sigma_{\rm 2halo}$ are correction terms that take into account, respectively, non-weak shear, cluster miscentering, and the contribution to the signal from large-scale structure; and $p_{\rm cc}$ is a free parameter related to the miscentering term, and represents the percentage of correctly centered clusters in each stack. 	
			
		Each term and the free parameters of the {\it Basic Model} are described in detail in the following sections.
							 
			As shown by \citet{gavazzi2007}, the two contributions to the shear signal from the luminous and dark matter can be distinguished by fitting a two-component mass model, which takes into account the contribution from the stellar mass of the halo central galaxy  $M_{\rm BCG}$. In order to model the BCG signal, we follow \citet{johnston2007} and add a point mass term to Equation \ref{deltasigmatot} (hereafter {\it Two Component Model}):\\

			\begin{multline}\label{deltasigmatot2}
				\Delta\Sigma(R)=\frac{M_{\rm BCG}}{\pi R^{2}} + \\p_{\rm cc}[ \Delta\Sigma_{\rm NFW}(R)+ \Delta\Sigma_{\rm nw}(R)]+ \\ (1-p_{\rm cc})\Delta\Sigma_{\rm sm}(R)+\Delta\Sigma_{\rm 2halo}(R)
			\end{multline}
			
			The BCG mass, $M_{\rm{BCG}}$, is either fixed at the value of the mean BCG stellar mass in each bin (hereafter $M_{\rm BCG}^{*}$), or left as a free parameter in the fit. We obtained $M_{\rm BCG}^{*}$ using our photometric and photometric redshift catalogs from \citet{raichoor2014}, and \citet{bruzual2003} stellar population models with \textit{LePhare}, in fixed redshift mode at the galaxy photometric redshift.
			
			 Previous works \citep{becker2007, rozo2009} have also shown that, when fitting the model profile to the halo profile derived from the observations in richness bins, the intrinsic scatter between the dark matter halo mass and the richness biases mass measurements. Following their modeling, we assume that the mass $M_{\rm 200}$ has a log-normal distribution at fixed richness, with the variance in $\ln{M_{\rm 200}}$, $\sigma_{\rm \ln{M200}|\lambda}$, and we add $\sigma_{\rm \ln{M200}|\lambda}$ to our {\it Basic Model} (hereafter {\it Added Scatter Model}).
			
			All the averages in the equations below are performed using the same weighting as in equation \ref{eq:mean1}.
		
		\subsubsection{$\Delta\Sigma_{\rm NFW}$ Profile}
			For the cluster halo profile, we assume an NFW profile. Numerical simulations have shown that dark matter halos density profiles, resulting from the dissipationless collapse of density fluctuations, can be well-described by this profile:
				
			\begin{subequations}\label{nfw}		
				\begin{align}
					\rho_{\rm NFW}(r)=\frac{\delta_{\rm c}\rho_{\rm c}}{(\frac{r}{r_{\rm s}})(1+\frac{r}{r_{\rm s}})^{2}} \tag{\ref{nfw}}\\
					\rho_{\rm c}=\frac{3H(z)^{2}}{8 \pi G}\\					
					r_{\rm s}=\frac{r_{\rm 200}}{c}\\	
					\delta_{\rm c}=\frac{200}{3}\frac{c^{3}}{\ln{(1+c)}-\frac{c}{1+c}}
				\end{align}
			\end{subequations}			
			where $\rho_{\rm c}$ is the critical density of the universe; $c$ is the concentration parameter;  $\delta_{\rm c}$ is a dimensionless parameter that depends only on the concentration; $r_{\rm s}$ is the scale radius of the cluster;  and $r_{\rm 200}$ is the radius at which the density is 200 times the critical density of the universe and can be considered as an approximation of the virial radius of the halo. The mass $M_{\rm 200}$ is the mass of a sphere of radius $r_{\rm 200}$ and average density of $200\rho_{\rm c}$:
				
			\begin{equation}\label{mass}
				M_{\rm 200}=M(r_{\rm 200})=\frac{4\pi}{3} r^{3}_{\rm 200} \times 200 \rho_{\rm c}
			\end{equation}
			
			Simulations have also shown that there is a relation between $M_{\rm 200}$ and $c$ \citep[e.g.][]{NFW1996, bullock2001}. In order to take this into account, we use the \citet{dutton2014} mass--concentration relation:
			 
			 \begin{equation}
			 	\log{c_{\rm 200}}=a+b\log{\left ( M_{\rm 200}/[10^{12}h^{-1}M_{\rm \odot}] \right )}
			 \end{equation}
			 
			 with $a=0.520+(0.905-0.520)\exp{(-0.617z^{1.21})}$ and $b=-0.101+0.026z$. This reduces the dimensionality of the model to one parameter, $r_{\rm 200}$, from which we can calculate the halo mass using Equation \ref{mass}.

			Integrating the tridimensional NFW density profile along the line of sight, we can calculate the NFW surface density:
			
			\begin{equation}
				\Sigma_{\rm NFW}(R)=2\int^{\infty}_{\rm 0}{\rho_{\rm NFW}(R,z)dz}
			\end{equation}
			Integrating again, we get $\overline{\Sigma}_{\rm NFW}(R)$, the average surface density inside a radius $R$:
			\begin{equation}
				\overline{\Sigma}_{\rm NFW}(<R)=\frac{2}{R^{2}}\int^{R}_{\rm 0}{R'\Sigma_{\rm NFW}(R')dR'}
			\end{equation}
			Finally, we can calculate the first term in Equation \ref{deltasigmatot}:\\
			 \begin{equation}
			 	\Delta\Sigma_{\rm NFW}=\overline{\Sigma}_{\rm NFW}(<R)-\Sigma_{\rm NFW}(R)
			\end{equation} 
	
		\subsubsection{Miscentering Term}
		
			Because the NFW density profile is spherically symmetric, an error in the determination of the halo center would lead to systematic underestimation of the lens mass. In fact, the random stacking offset smooths the differential surface mass density profile \citep{george2012}.

			Following \citet{licitra2016a}, we use both simulations and X-ray observations to obtain a model of the distribution of the offsets between the RedGOLD center and the cluster true center. We apply RedGOLD to the lightcones of \citet{henriques2012}, and calculate the offsets between the centers estimated by the algorithm and the true centers from the simulations. We also match our RedGOLD detections to X-ray detections in the same areas  \citep{gozaliasl2014} to measure our average offset between RedGOLD and X-ray cluster centers. We perform the match between the RedGOLD and the  \citet{gozaliasl2014}  catalogs by imposing a maximum separation between centers of $1~Mpc$ and a maximum difference in redshift of $\Delta z =0.1$. 
						
			\begin{figure*}[!htbp]
                        \centering
                        \includegraphics[scale=.50]{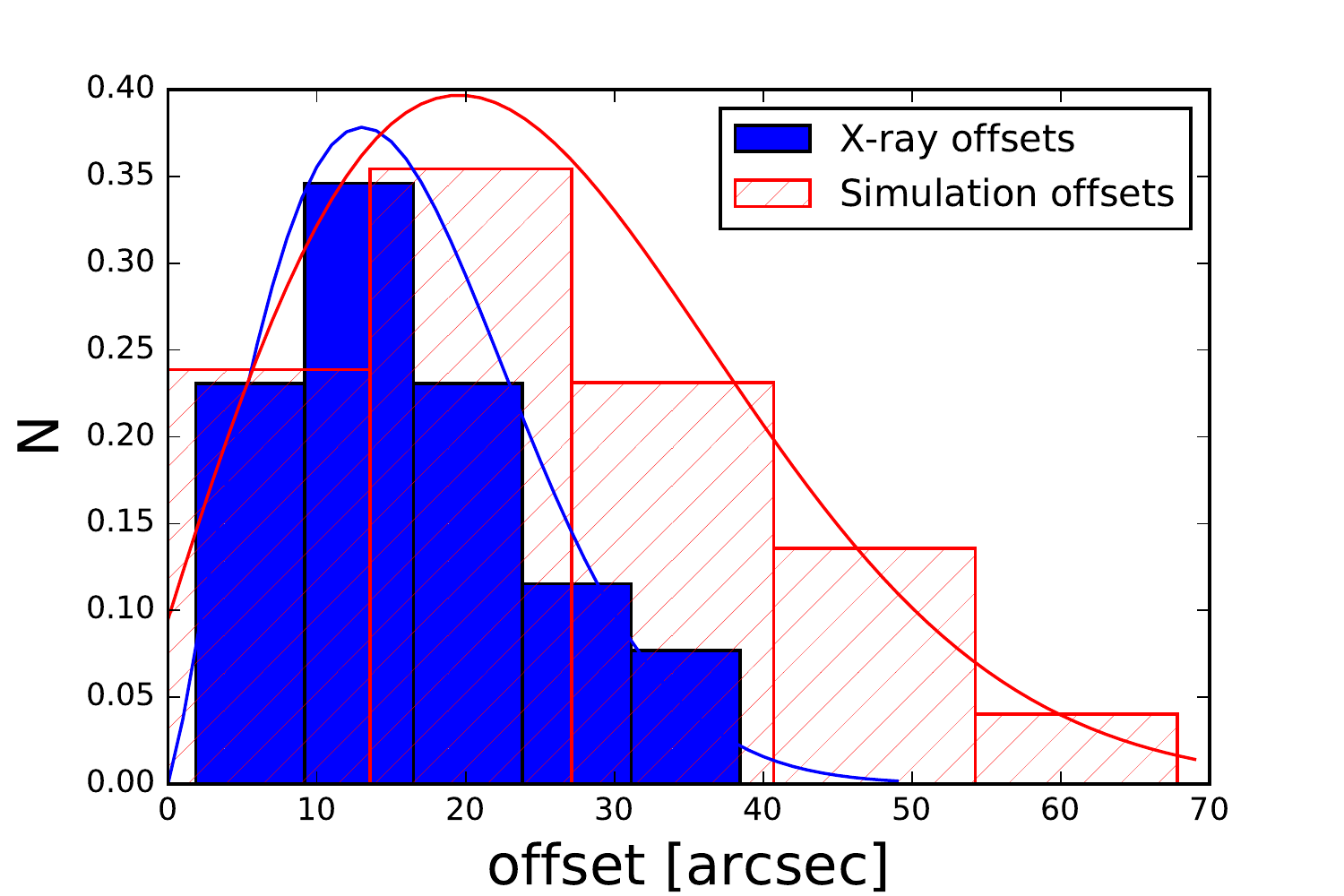}
                        \includegraphics[scale=.50]{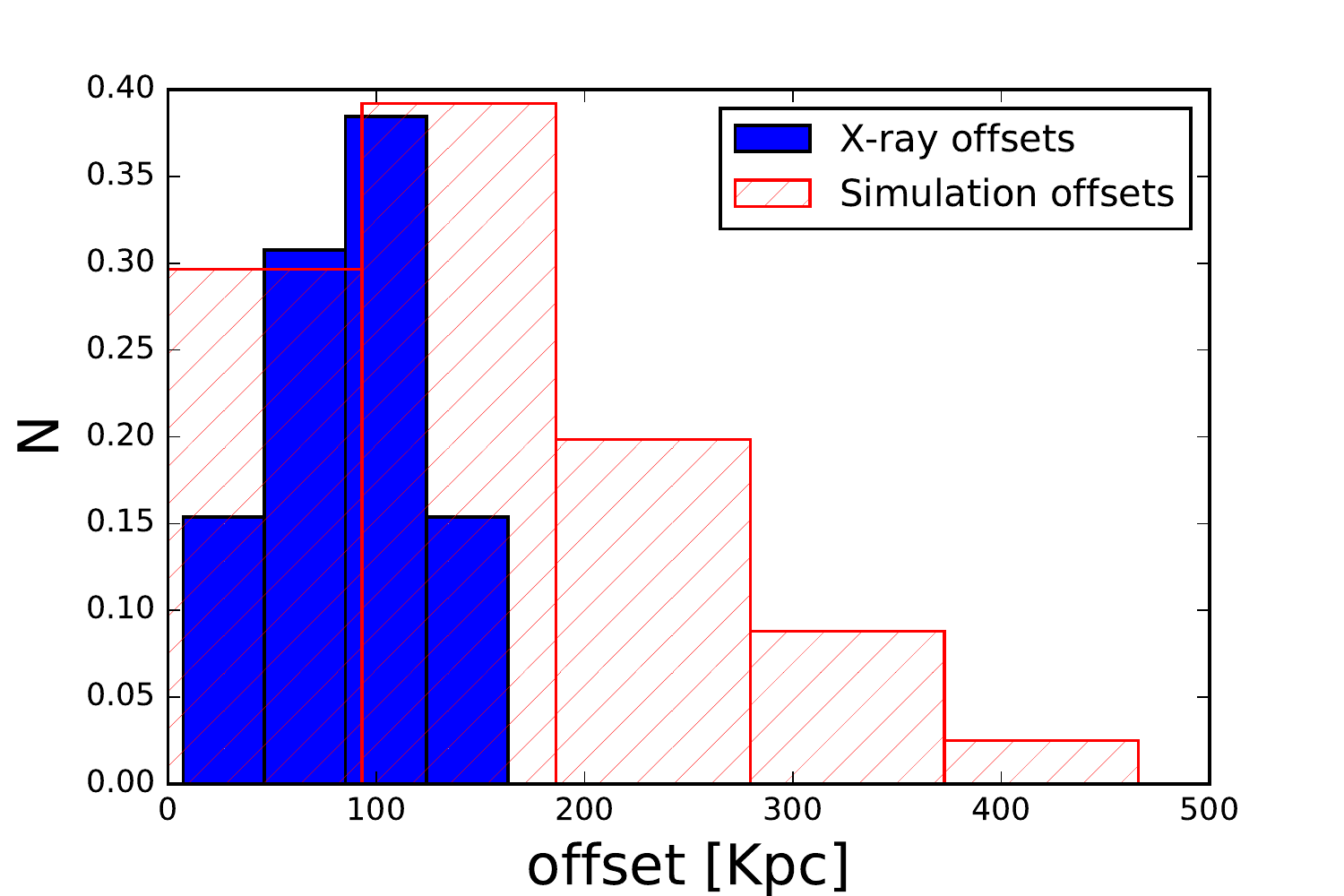}
                        \caption{\scriptsize On the left, distribution of the offsets, in arcsec, between the RedGOLD and X-ray cluster centers (in blue), and between the RedGOLD and the \citet{henriques2012} simulation centers (in red). The blue and red lines show the fitted Rayleigh distributions with modes of 13 and 23 arcsec, respectively. On the right, the offset distributions in $kpc$.}
                        \label{fig:offset}
                        \end{figure*}
			
			In both cases, we find that the distribution of the offsets on the plane perpendicular to the line of sight can be modeled as a Rayleigh distributions with modes of 23 and $13~arcsec$, respectively \citep[Figure \ref{fig:offset}, on the left; see also][]{johnston2007, george2012, ford2015}. What is important is that the model (a Rayleigh distribution) is the same in both cases, even if the precise values of the mode are different. In fact, the mode of the Rayleigh distribution, from which its mean, median and variance can be derived,  will be derived as a free parameter from our analysis. In Figure 2, on the right, we also show the offset distributions in kpc. A Rayleigh distribution is also consistent with the published center offset distribution predicted from cosmological simulations for X-ray detected clusters, including AGN feedback \citep{cui2016}.
			
			We assume that this distribution represents the general offset distribution for our entire RedGOLD sample $P(R_{\rm off})$, and model it following \citet{johnston2007}:
			\begin{equation}
				P(R_{\rm off})=\frac{R_{\rm off}}{\sigma^{2}_{\rm off}}\exp \left[-\frac{1}{2}\left(\frac{R_{\rm off}}{\sigma_{\rm off}}\right)^{2}\right]
			\end{equation}
			where $R_{\rm off}$ is the offset between the true and the estimated center, projected on the lens plane,  and $\sigma_{\rm off}$ is the mode, or scale length, of the distribution. The surface density measured at the coordinates $(R, \theta)$, with $\theta$ the azimuthal angle, relative to the offset position, $R_{\rm off}$, is:
			\begin{multline}
				\Sigma_{\rm NFW}(R,\theta | R_{\rm off})=\\ \Sigma_{\rm NFW}\left(\sqrt{R^{2}+R^{2}_{\rm off}-2RR_{\rm off}\cos \theta}\right)
			\end{multline}
			and the azimuthal averaged surface density around $R_{\rm off}$ is given by:
			\begin{equation}
				\Sigma_{\rm NFW}(R|R_{\rm off})=\frac{1}{2\pi}\int^{2\pi}_{\rm 0}{\Sigma_{\rm NFW}(R,\theta | R_{\rm off})d\theta}
			\end{equation}
			
			To model the effect of miscentering, we smooth the $\Sigma_{\rm NFW}(R|R_{\rm off})$ profile convolving it with $P(R_{\rm off})$:
			\begin{equation}
				\Sigma_{\rm sm}(R)=\int^{\infty}_{\rm 0}{\Sigma_{\rm NFW}(R|R_{\rm off})P(R_{\rm off})dR_{\rm off}}
			\end{equation}
			and obtain the stacked surface density profile $\Sigma_{\rm sm}(R)$ around the offset positions of our ensemble of clusters with offset distribution $P(R_{\rm off})$ \citep{yang2006, johnston2007, george2012}.
			
			Finally we can write the miscentering term as:
			\begin{equation}
				\Delta\Sigma_{\rm sm}(R)=\overline{\Sigma}_{\rm sm}(<R)-\Sigma_{\rm sm}(R)
			\end{equation}
			with $\overline{\Sigma}_{\rm sm}(<R)$ being, as before, the average surface density within the radius $R$.
			
			The miscentering term adds two free parameters to our model, $\sigma_{\rm off}$ and $p_{\rm cc}$, which is the percentage of correctly centered clusters in the stack, already introduced in Equation \ref{deltasigmatot}.

		\subsubsection{Non-weak Shear Term}
			The non-weak shear correction arises from the fact that what we actually measure is the reduced shear:
			\begin{equation}
				g_{\rm t}=\frac{\gamma_{\rm t}}{1-k}
			\end{equation}
			where $k\equiv\Sigma_{\rm NFW}/\Sigma_{\rm c}$ is the convergence. Usually in the weak lensing regime $g_{\rm t} \approx \gamma_{\rm t}$, if $\gamma_{\rm t} << 1$ and $k<<1$, but for relatively massive halos, this assumption may no longer hold at the innermost radial bins in which we want to measure the cluster profile.
			
			As described in \citet{johnston2007}, we introduce the non-weak shear correction term, calculated in \citet{mandelbaum2006}. In the non-weak regime, the tangential ellipticity component, $\epsilon_{\rm t}$ is proportional to $g_{\rm t}$, instead of $\gamma_{\rm t}$. We can expand $\epsilon_{\rm t}$ in power series as: 	
			\begin{multline}\label{eq:nonlin}
				\epsilon_{\rm t}=\sum_{\rm n=0}^{\infty}Ag_{\rm t}^{2n+1}\\=A\left( \frac{\gamma_{\rm t}}{1-k}\right)^{2n+1}=A\left( \frac{\Delta\Sigma\Sigma_{\rm c}^{-1}}{1-\Sigma\Sigma_{\rm c}^{-1}}\right)^{2n+1}
			\end{multline}		
			As shown in detail in appendix A of \citet{mandelbaum2006}, we can calculate the correction term from the expansion in power series to the second order of $\epsilon_{\rm t}$, in powers of $\Sigma_{\rm c}$. We obtain the following term, which we add in Equation \ref{deltasigmatot}:					
			\begin{equation}
				\Delta\Sigma_{\rm nw}(R)=\Delta\Sigma_{\rm NFW}(R)\Sigma_{\rm NFW}(R)\frac{\left<\Sigma^{-3}_{\rm c}\right>}{\left<\Sigma^{-2}_{\rm c}\right>}
			\end{equation}
			
			\begin{table}[!htbp]
\begin{center}
\resizebox{.48\textwidth}{!}{
\begin{tabular}{L{2cm}C{1cm}C{2cm}C{5.5cm}}
\tableline\tableline\noalign{\smallskip}
& Basic Model & Added Scatter Model & Two Component Model\\[2pt]
\tableline
\tableline\noalign{\smallskip}
$r_{\rm 200} (\rm Mpc)$ & (0, 2) & --- & (0, 2)\\
$\sigma_{\rm off}{(\rm arcmin)}$ & (0, 2) & (0, 2) & (0, 2)\\
$p_{\rm{cc}}$ & (0, 1) & (0, 1) & (0, 1)\\
$\sigma_{\rm \ln{M}|\lambda}$ & --- & (0.1, 0.7) & ---\\
$\log{\rm (M_{200}/M_{\odot})}$ & --- & (11, 17) & ---\\
$\log{\rm (M_{BCG}/M_{\odot})}$& --- & --- & (9, 13)~or~fixed~at~$\log{(M_{\rm BCG}^{*}/M_{\odot})}$\\
\tableline
\end{tabular}
}
\caption{\scriptsize MCMC uniform prior ranges for the different parameters of the three Models. The lack of a numerical value indicates that the parameter is not included in the respective model.}
\label{tab:priors}
\end{center}
\end{table}

		\subsubsection{Two-halo Term}
			On large scales, the lensing signal is dominated by nearby mass concentrations, halos, and filaments. \citet{seljak2000} developed an analytic halo model in which all the matter in the universe is hosted in virialized halos described by a universal density profile. They computed analytically the power spectrum of dark matter and galaxies, and their cross-correlation based on the \citet{press1974} model. They found that, ignoring the contribution from satellite galaxies, a cluster can be modeled by two contributions: the one-halo term and the two-halo term. The first represents the correlation between the central galaxy and the host dark matter halo and corresponds to $\Delta\Sigma_{\rm NFW}(R)$. The second accounts for the correlation between the cluster central galaxy and the host dark matter halo of another cluster. 
			
			On large scales, the two-halo power spectrum is proportional to the halo bias and the linear power spectrum, $P_{\rm 2halo}\propto b(M_{\rm 200, z})P_{\rm lin}(k)$. In order to calculate the surface density associated to the two-halo term, we integrate the galaxy-dark matter linear cross-correlation function $\xi_{\rm lin}(r)$, obtained by the Fourier transform of the linear power spectrum.
			
			Following  \citet{johnston2007} and \citet{ford2015}, we can write the two-halo term as:		
			\begin{multline}\label{bias}
				\Delta\Sigma_{\rm 2halo}(R,b)= \\ b(M_{\rm 200},z)\Omega_{\rm m}\sigma^{2}_{\rm 8}D(z)^{2}\Delta\Sigma_{\rm l}(R)
			\end{multline}			
			where $b(M_{\rm 200},z)$ is the bias factor, $\Omega_{\rm m}$ is the matter density parameter, $\sigma^{2}_{\rm 8}$ is the amplitude of the power spectrum on scales of 8 $h^{-1}Mpc$, $D(z)$ is the growth factor and
			
			\begin{equation}
				\Delta\Sigma_{\rm l}(R)=\overline{\Sigma}_{\rm l}(<R)-\Sigma_{\rm l}(R)
			\end{equation}
			where
			\begin{multline}
				\Sigma_{\rm l}(R,z)= \\ (1+z)^{3}\rho_{\rm c,0}\int^{\infty}_{\rm -\infty}{\xi_{\rm lin}\left((1+z)\sqrt{R^{2}+y^{2}}\right)}dy
			\end{multline}
			
	The factor $(1+z)$ arises from the conversion from physical units to comoving units. 
			
			 For the bias factor, we use the analytic formula calculated by \citet{seljak2004}, and for $P_{\rm lin}(k)$, we use tabulated values from CAMB \citep{lewis2000}.

		\section{RESULTS}
			\subsection{Cluster Mass Estimation}
			
				\subsubsection{Fit the Profile Model to the Shear Profile} \label{subsec:fit}
					
				We fit the shear profiles, obtained as described in Section \ref{measured_prof} with the density profile models of Section \ref{model_prof}, progressively adding model parameters to quantify their impact on the final results.
				
			We start from the {\it Basic Model} with an NFW surface density contrast and correction terms that take into account cluster miscentering, non-weak shear, and the two halo term. This model has three free parameters: the radius $r_{\rm 200}$, from which we calculate the mass $M_{\rm 200}$ from Equation \ref{mass}, and the miscentering parameters $p_{\rm cc}$, and $\sigma_{\rm off}$.
			
					We then take into account the intrinsic scatter in the mass--richness relation through the {\it Added Scatter Model}, which has four free parameters: $\log{M_{200}}$, $p_{\rm cc}$, $\sigma_{\rm off}$, and $\sigma_{M|\lambda}$. For each bin, we use the mass--richness relation, calculated from the {\it Basic Model} to infer the mean mass of the stacked clusters, as a first approximation. We then randomly scatter the mass using a gaussian distribution with mean $\left<\ln{M_{\rm 200}}\right>$ and width $\sigma_{\rm \ln{M200}|\lambda}$.
					
					Finally, we consider the {\it Two Component Model}, with four free parameters: $r_{\rm 200}$, $p_{\rm cc}$, $\sigma_{\rm off}$, and $\log{M_{\rm BCG}}$. When we fix the BCG mass to the mean stellar mass for each richness bin, $M_{\rm BCG}=M_{\rm BCG}^{*}$, the free parameters reduce to three.
					
					The parameters used in each case are summarized in Table~\ref{tab:priors}.
				
				We perform the fit using Markov Chains Monte Carlo \citep[MCMC;][]{metropolis1953}. This method is particularly useful when the fitting model has a large number of parameters, the posterior distribution of the parameters is unknown, or the calculation is computationally expensive. MCMC allows efficient sampling of the model likelihood by constructing a Markov chain that has the target posterior probability distribution as its stationary distribution. Each step of the chain is drawn from a model distribution and is accepted or rejected based on the criteria defined by the sampler algorithm. 
				
								\begin{figure*}[!htbp]
\centering
\includegraphics[scale=.80]{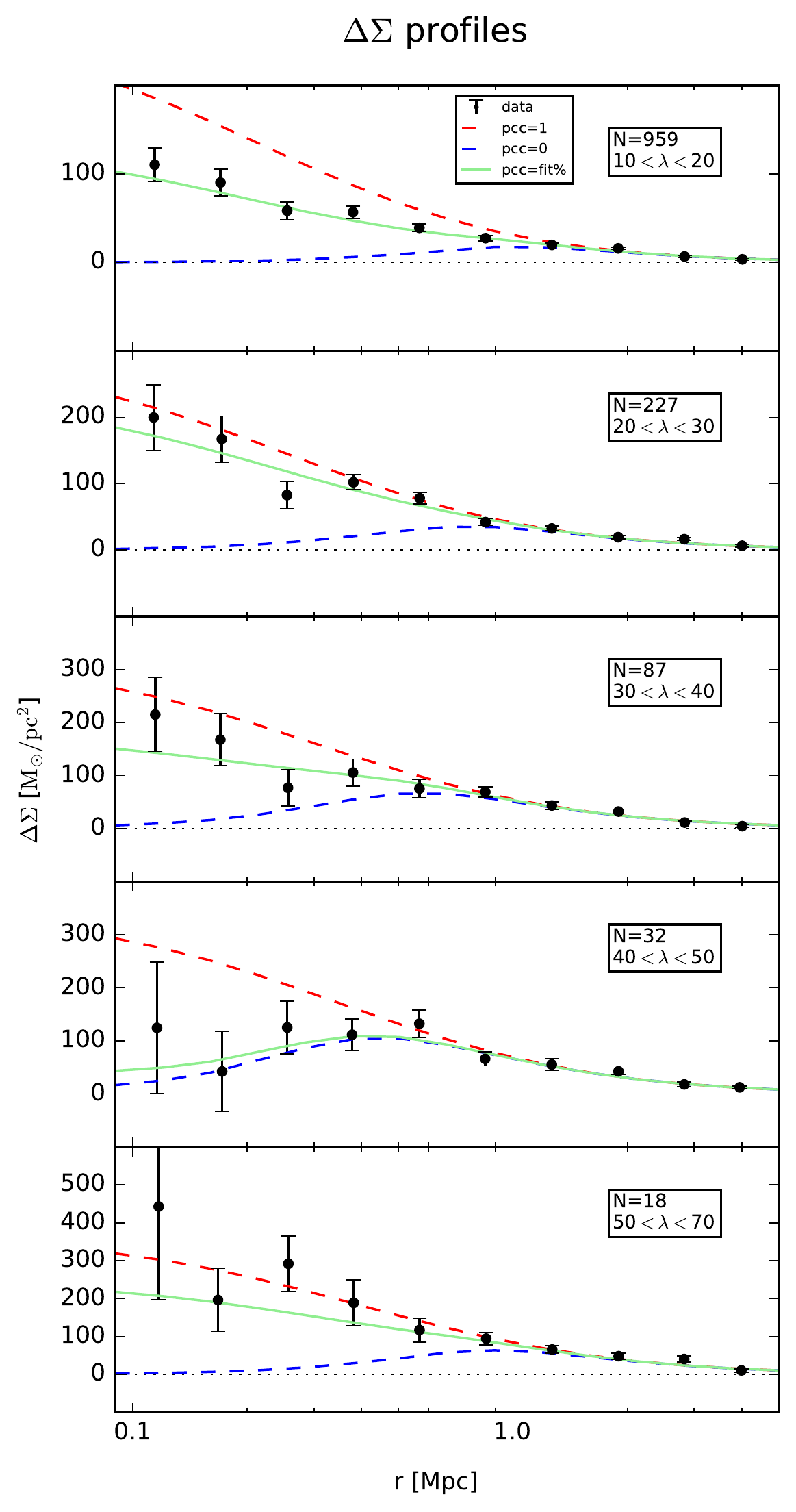}
\raisebox{-0.08\height}{\includegraphics[scale=.40]{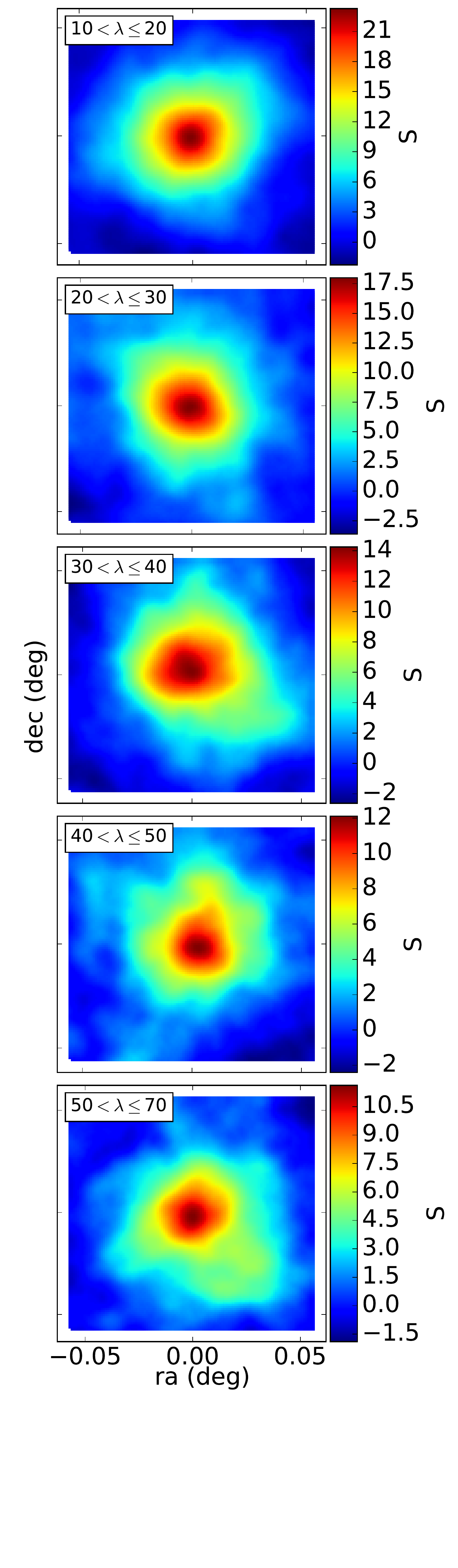}}
\vspace*{-15mm}
\caption{\scriptsize On the left: shear profiles measured with our weak lensing analysis selected sample (CFHT-LS W1 + NGVS5+ NGVS4). The fits were  obtained using  the {\it Basic Model}. We show our shear profile measurement (black), the fit results (green), the ideal profiles that we would obtain in the case in which all the clusters in the stack were perfectly centered (red) and when they would have been all miscentered (blue). The dotted lines show $\Delta\Sigma(R)=0$. We get similar results using the Added Scatter and the Two Component Models. On the right: lensing signal-to-noise ratio maps in each richness bin for our weak lensing selected sample. We applied aperture mass statistics.}
\label{fig:profiles}
\end{figure*}
				
				To run our MCMC, we use \textit{emcee}\footnote{https://github.com/dfm/emcee} \citep{emcee}, a Python implementation of the parallel Stretch Move by \citet{goodman2010}. In order to choose the starting values of the chain we first perform a minimization with the Python version of the Nelder--Mead algorithm, also known as downhill simplex \citep{nelder1965}. We used flat priors  (i.e. a uniform distribution within a given range) for all parameters. Our initial priors, for the three different models, are shown in Table \ref{tab:priors}. All parameters are constrained to be positive and inside a range chosen according to their physical meaning.  To choose the range for the intrinsic scatter, we refer to the values calculated by \citet{licitra2016a}. They found $\sigma_{\rm \ln{M}|\lambda}=0.39\pm0.07$ using the X-ray catalog of \citet{gozaliasl2014} and $\sigma_{\rm \ln{M}|\lambda}=0.30\pm0.13$ from \citet{mehrtens2012}.
				
				MCMC produce a representative sampling of the likelihood distribution, from which we obtain the estimation of the error bars on the fitting parameters and of the confidence regions for each couple of parameters. We calculate the model likelihood using the bootstrap covariance matrix of Equation \ref{eq:cov}:
				\begin{multline}
					ln\mathcal{L}=-\frac{1}{2}\left(\Delta\Sigma_{\rm data}- \Delta\Sigma_{\rm model}\right)^{T}C^{-1}\\ \left(\Delta\Sigma_{\rm data}- \Delta\Sigma_{\rm model} \right)
				\end{multline}	
				We use an ensemble of 100 walkers, a chain length of 1000 steps and a burn-in of 100 steps leading to a total of 90,000 points in the parameters space. In order to test the result of our chain, we check the acceptance fraction and the autocorrelation time, to be sure that we efficiently sample the posterior distribution and have enough independent samples.

			\subsubsection{Fit Parameters}\label{results}
			
				We perform the fit of the models to the observed profiles on each of the three samples, CFHT-LS W1, NGVS5, and NGVS4.  We then combine the CFHTLS and NGVS5, and all the three samples together. 
					
				The profiles obtained using the {\it Basic Model} and the complete sample (CFHT-LS W1 + NGVS5 + NGVS4) are shown in black in Figure \ref{fig:profiles}, on the left. The error bars on the shear profiles are the square root of the diagonal elements of the covariance matrix.  
				
				\begin{figure}[!htbp]
\centering
\includegraphics[scale=.60]{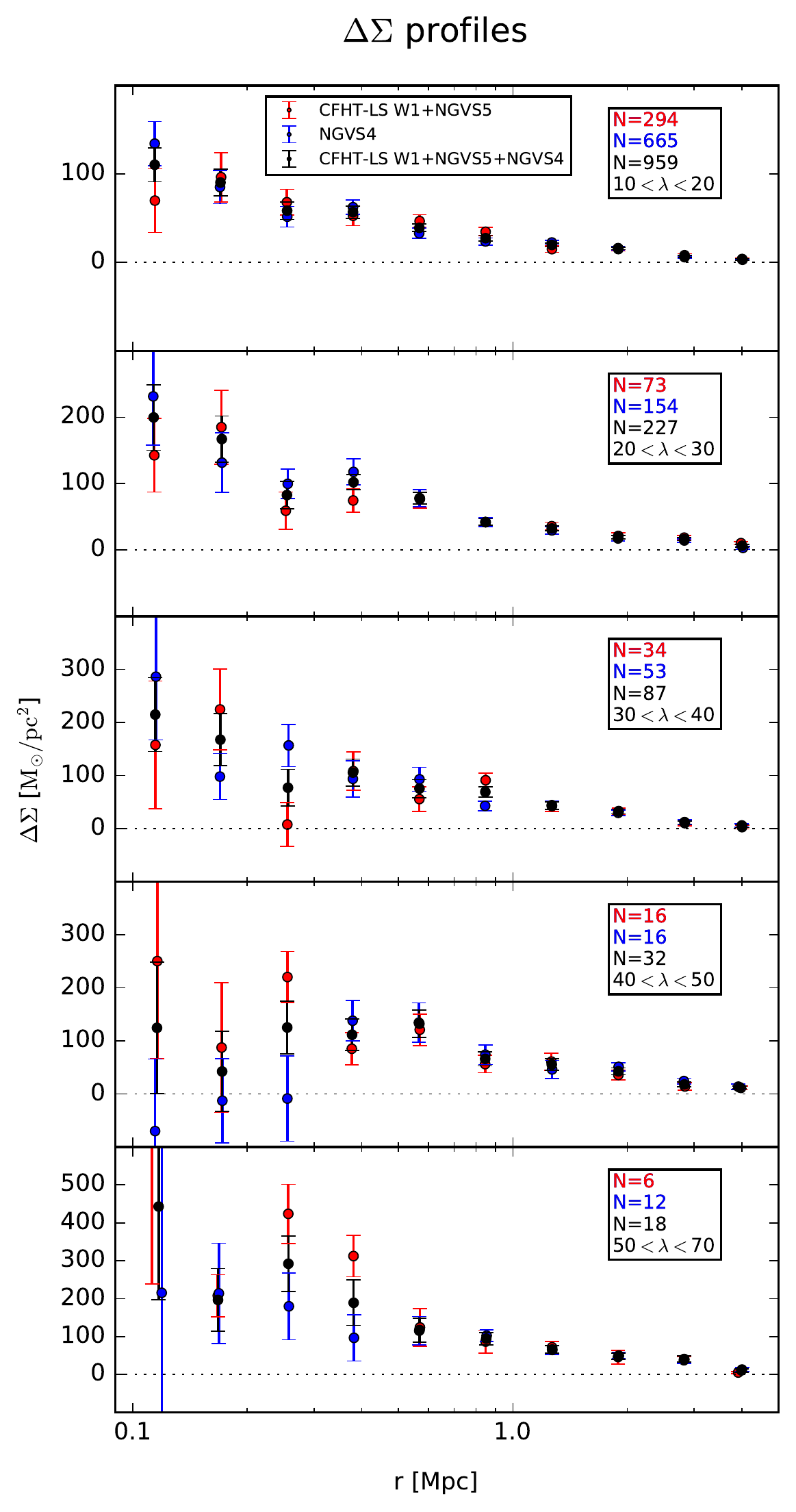}
\caption{\scriptsize Shear profiles measured with the weak lensing {\it selected} CFHT-LS W1 + NGVS5, in red,  with weak lensing selected NGVS4, in blue, and with the weak lensing {\it selected} CFHT-LS W1 + NGVS5 + NGVS4, in black. We notice that the addition of the four bands sample does not significantly change the profiles. The profiles measured using the three different samples are compatible within $1\sigma$ and the profiles obtained using CFHT-LS W1 + NGVS5 + NGVS4 have smaller error bars.}
\label{fig:profiles_comp}
\end{figure}

				 The profiles measured using the CFHT-LS W1 + NGVS5 sample, the NGVS4 sample, and the complete sample are shown in Figure \ref{fig:profiles_comp}. They are consistent within $1\sigma$ and the error bars are smaller in the last case. We can conclude that the richness shifts applied to NGVS4 seem not to bias our results when this sample is added to the other two that are covered by five bands. Increasing the sample size, we notice a progressive improvement in the profiles that are recovered with a lower noise level.	
							
				Because the miscentering correction is the one that most affects the mass estimation, in Figure \ref{fig:profiles}, on the left, we show the fitted profiles (green lines), and the profiles that we would obtain with and without the miscentering term. The red lines represent the profiles we would obtain in the case in which all the clusters in the stack were perfectly centered ($p_{\rm cc}=1$), and the blue lines show the opposite case ($p_{\rm cc}=0$). An incorrect modeling of this effect leads to biased mass values \citep[i.e. mass underestimation between 10 and $40\%$,][]{ford2015}.
				
				In Figure \ref{fig:profiles}, on the right, we show the lensing S/N maps. These maps were calculated using aperture mass statistics \citep{schneider1996, schirmer2006, dufan2014}. For each richness bin, we create a grid with a side of $1~Mpc$ and binning of $0.001~deg$, centered on the stacked clusters. In each cell, we evaluate the amount of tangential shear, filtered by a function that maximizes the S/N of an NFW profile, inside a circular aperture, following \citet{schirmer2006}. For stacked clusters, a $S/N\sim10$ is considered sufficient to recover the fitting parameters \citep{oguri_takada2011}. All richness bins have $S/N\geq10$. The highest richness bin shows the lowest S/N, being less populated than the others.

				We show the results of our fits in Table \ref{tab:fit}, for the {\it Basic Model}, {\it Added Scatter Model}, and {\it Two Component Model}. The values of the radius, of the mass, and of the miscentering parameters for each richness bin are consistent within 1$\sigma$ for the three models. We found that the intrinsic scatter and BCG mass are not constrained by the data. The main effect of the addition of $\sigma_{M|\lambda}$ to the fit is to introduce more uncertainties and to increase the error on the estimated parameters. The inclusion of $M_{\rm BCG}$ in the model (either set as a free parameter or fixed to $M_{\rm BCG}^{*}$) has no impact on the estimated parameters, which are therefore the same as those obtained using the {\it Basic Model}. We can conclude that the contribution of the BCG mass is not significant in the radial range we are using to fit the shear profiles.  
																	
\begin{figure*}[!htbp]
\centering
\includegraphics[scale=.70]{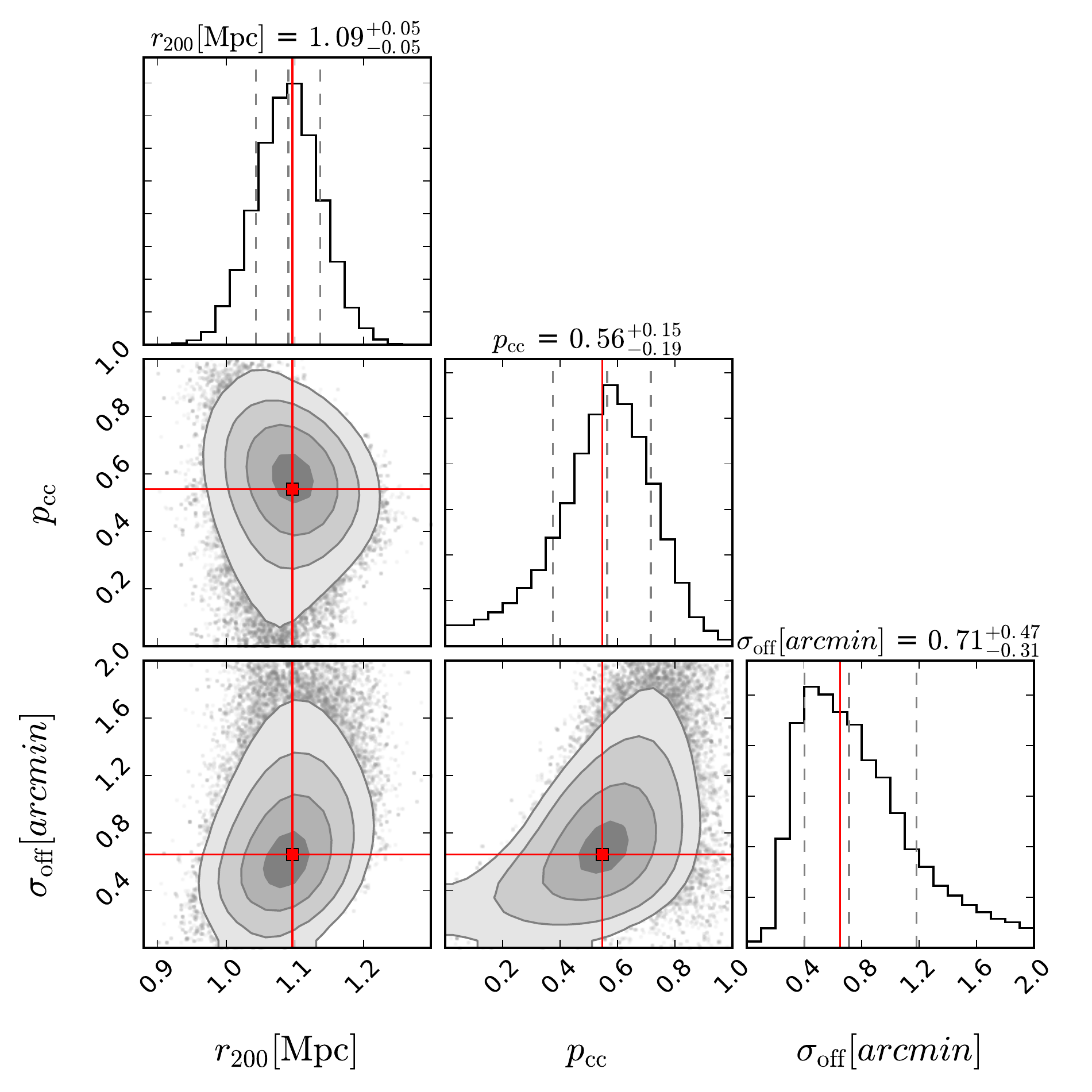}
\caption{\scriptsize Confidence levels on the fit parameters obtained with our MCMC analysis, for the third richness bin, fitted with the Two Component Model. We obtained a similar plot for each richness bin and model. On the diagonal, we show the 1D histograms of each parameter. The 2D histograms are also shown for each couple of parameters with confidence levels corresponding to $0.5\sigma$, $1\sigma$, $1.5\sigma$, and $2\sigma$. The parameter values and errors are based on the 16th, 50th, and 84th quantiles (shown as dashed lines in the 1D histograms). The red squares and lines represents the values that correspond to the maximum likelihood. We obtained similar plots for each richness bin.}
\label{fig:corner}
\end{figure*}

				In Figure \ref{fig:corner}, we show an example of error bars and the confidence regions of the parameters, obtained using the python package \textit{corner} by \citet{corner}. This example corresponds to the third richness bin, fitted with the {\it Two Component Model}. On the diagonal, we show the one-dimensional histograms of the parameter values, representing the marginalized posterior probability distributions. Under the diagonal, we show the two-dimensional histograms for each couple of parameters and the confidence levels corresponding to $0.5\sigma$, $1\sigma$, $1.5\sigma$ and $2\sigma$.

			\subsection{Mass--Richness Relation}
				Using the mass measured for each richness bin, we perform a fit to a power law to infer the mass--richness relation for all three models, using the python orthogonal distance regression routine \citep{boggs1990} to take into account the errors in both  $\log{\lambda}$ and $\log{M_{\rm 200}}$:
				\begin{equation}
					\log{M_{\rm 200}}=\log{M_{\rm 0}}+\alpha\log{\lambda/\lambda_{\rm 0}}
				\end{equation}
				with a pivot richness $\lambda_{\rm 0}=40$. 
												
\begin{figure}[!htbp]
\centering
\includegraphics[scale=.50]{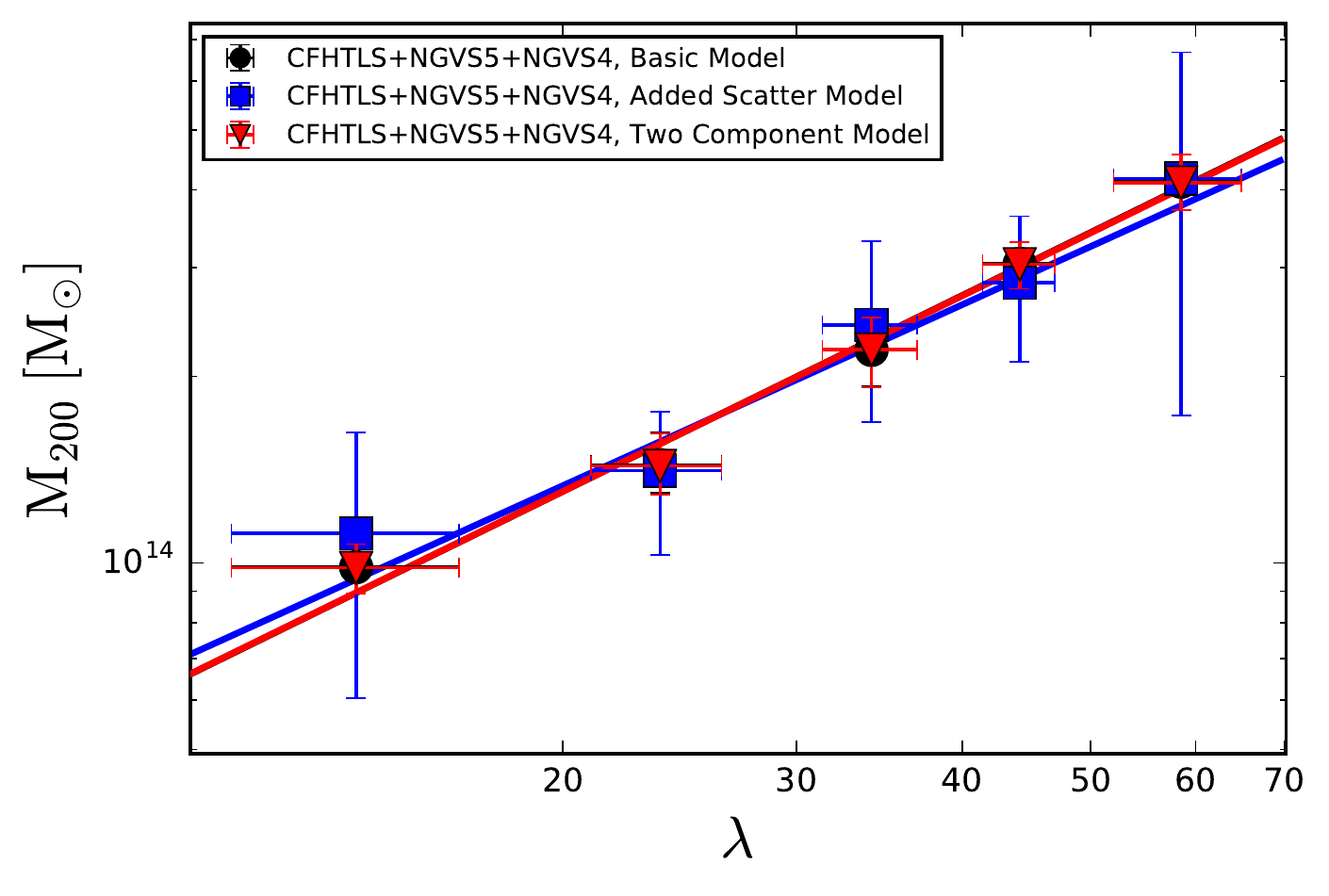}
\caption{\scriptsize The weak lensing mass--richness relations obtained with the weak lensing {\it selected} CFHT-LS W1 + NGVS5 + NGVS4, using the {\it Basic Model} (black line and black dots), Added Scatter Model (blue line and blue squares), Two Component Model (red line and red triangles). See text for the description of the models.}
\label{fig:massrich}
\end{figure}

				In Table \ref{tab:mr_par} and in Figure \ref{fig:massrich}, we show the results obtained fitting the three models. The slope and the normalization values are all consistent within $1 \sigma$, for the three models. We notice that the uncertainties in the fit of the {\it Added Scatter Model} are larger, due to the inclusion of the intrinsic scatter as a free parameter.

				\begin{table*}[!htbp]
\begin{center}
\resizebox{.8\textwidth}{!}{
\begin{tabular}{lllllllllll}
\tableline\tableline\noalign{\smallskip}
$\lambda$ Range & N & $\lambda$ & z & Model & $r_{\rm 200}$ & $M_{\rm 200} $ &  $\sigma_{\rm off}$  & $p_{\rm{cc}}$ & $\sigma_{\rm \ln{M}|\lambda}$ & $M_{\rm BCG}~(M_{\rm BCG}^{*})$\\
&&&&&Mpc&$10^{13}M_{\rm \odot}$&arcmin& & &$10^{11}M_{\rm \odot}$\\[2pt]
\tableline
\tableline\noalign{\smallskip}
				&	&		    &					& Basic & $0.83^{+0.03}_{\rm -0.03}$ & $10^{+1}_{\rm -1}$ & $1.5^{+0.3}_{\rm -0.3}$ & $0.5^{+0.1}_{\rm -0.1}$ & -- & --\\[5pt]
$10<\lambda\le20$ & 959 & $14\pm3$ & 0.40\multirow{3}{*}{}   & Added Scatter & $0.86^{+0.13}_{\rm -0.13}$ & $11^{+5}_{\rm -5}$ & $1.5^{+0.3}_{\rm -0.3}$ & $0.5^{+0.1}_{\rm -0.1}$ & $0.4^{\rm +0.2}_{\rm -0.2}$ & --\\[5pt]
				&	&		    &					& Two Component & $0.83^{+0.02}_{\rm -0.03}$ & $10^{+1}_{\rm -1}$ & $1.5^{+0.3}_{\rm -0.3}$ & $0.5^{+0.1}_{\rm -0.1}$ & -- & $1^{\rm +3}_{\rm -3}~(1.53^{\rm +0.02}_{\rm -0.02})$\\[5pt]
\tableline\noalign{\smallskip}
				&	&		    &					& Basic & $0.94^{+0.03}_{\rm -0.04}$ & $14^{+1}_{\rm -2}$ & $1.0^{+0.7}_{\rm -0.7}$ & $0.7^{+0.2}_{\rm -0.1}$ & -- & --\\[5pt]
$20<\lambda\le30$ & 227 & $24\pm3$ & 0.39\multirow{3}{*}{}   & Added Scatter  & $0.94^{+0.08}_{\rm -0.08}$ & $14^{+4}_{\rm -3}$ & $0.9^{+0.7}_{\rm -0.8}$ & $0.8^{+0.2}_{\rm -0.1}$ & $0.4^{\rm +0.2}_{\rm -0.2}$ & --\\[5pt]
				&	&		    &					& Two Component & $0.94^{+0.03}_{\rm -0.04}$ & $14^{+1}_{\rm -2}$ & $1.0^{+0.7}_{\rm -0.7}$ & $0.7^{+0.2}_{\rm -0.1}$ & -- & $1^{\rm +3}_{\rm -3}~(1.7^{\rm +0.1}_{\rm -0.1})$\\[5pt]
\tableline\noalign{\smallskip}
				&	&		    &					& Basic & $1.09^{+0.05}_{\rm -0.05}$ & $22^{+3}_{\rm -3}$ & $0.7^{+0.3}_{\rm -0.5}$ & $0.6^{+0.2}_{\rm -0.1}$ & -- & --\\[5pt]
$30<\lambda\le40$ & 87 & $34\pm3$ & 0.39\multirow{3}{*}{}     & Added Scatter  & $1.12^{+0.11}_{\rm -0.13}$ & $24^{+7}_{\rm -9}$ & $0.7^{+0.3}_{\rm -0.4}$ & $0.6^{+0.2}_{\rm -0.2}$ & $0.4^{\rm +0.2}_{\rm -0.2}$ & --\\[5pt]
				&	&		    &					& Two Component & $1.09^{+0.05}_{\rm -0.05}$ & $22^{+3}_{\rm -3}$ & $0.7^{+0.3}_{\rm -0.5}$ & $0.6^{+0.2}_{\rm -0.2}$ & -- & $1^{\rm +3}_{\rm -3}~(1.8^{\rm +0.1}_{\rm -0.1})$\\[5pt]
\tableline\noalign{\smallskip}
				&	&		    &					& Basic & $1.21^{+0.04}_{\rm -0.03}$ & $30^{+3}_{\rm -3}$ & $0.5^{+0.1}_{\rm -0.1}$ & $0.1^{+0.1}_{\rm -0.1}$ & -- & --\\[5pt]
$40<\lambda\le50$ & 32 & $44\pm3$ & 0.39\multirow{3}{*}{}     & Added Scatter  & $1.18^{+0.10}_{\rm -0.11}$ & $28^{+7}_{\rm -8}$ & $0.5^{+0.1}_{\rm -0.1}$ & $0.1^{+0.1}_{\rm -0.1}$ & $0.4^{\rm +0.2}_{\rm -0.2}$& --\\[5pt]
				&	&		    &					& Two Component & $1.21^{+0.04}_{\rm -0.03}$ & $30^{+3}_{\rm -3}$ & $0.5^{+0.1}_{\rm -0.1}$ & $0.1^{+0.1}_{\rm -0.1}$ & -- & $1^{\rm +3}_{\rm -3}~(1.9^{\rm +0.2}_{\rm -0.2})$\\[5pt]
\tableline\noalign{\smallskip}
				&	&		    &					& Basic & $1.35^{+0.04}_{\rm -0.05}$ & $41^{+4}_{\rm -4}$ & $1.1^{+0.8}_{\rm -0.6}$ & $0.7^{+0.2}_{\rm -0.2}$ & -- & --\\[5pt]
$50<\lambda\le70$ & 18 & $59\pm6$ & 0.38\multirow{3}{*}{}     & Added Scatter  & $1.35^{+0.26}_{\rm -0.27}$ & $41^{+2}_{\rm -2}$ & $0.9^{+0.7}_{\rm -0.7}$ & $0.7^{+0.3}_{\rm -0.2}$ & $0.4^{\rm +0.2}_{\rm -0.2}$ & --\\[5pt]
				&	&		    &					& Two Component & $1.35^{+0.04}_{\rm -0.05}$ & $41^{+4}_{\rm -5}$ & $1.1^{+0.8}_{\rm -0.6}$ & $0.7^{+0.2}_{\rm -0.2}$ & -- & $1^{\rm +4}_{\rm -4}~(2.0^{\rm +0.2}_{\rm -0.2})$\\[5pt]
\tableline
\end{tabular}
}
\caption{\scriptsize Parameters derived from the fit of the Basic Model, Added Scatter Model, and Two Component Model shear profiles to our measurements. Here, $\lambda$ is the cluster optical richness derived with RedGOLD and the first column gives the richness range; $N$ is the number of stacked clusters in each bin; $z$ is the mean redshift; $r_{\rm 200}$ is the mean radius in Mpc; $M_{\rm 200}$ is the mean mass in units of $10^{13}M_{\rm \odot}$; $\sigma_{\rm off}$ is the scale length of the offset distribution in arcmin; $p_{\rm cc}$ is the percentage of correctly centered clusters in the stack; $\sigma_{\rm \ln{M}|\lambda}$ is the intrinsic scatter in the mass--richness relation; $M_{\rm BCG}~(M_{\rm BCG}^{*})$ is the mean BCG mass in units of $10^{11}M_{\rm \odot}$, left as a free parameter in the fit, and fixed at the stellar mass value recovered with \textit{LePhare}, respectively.}
\label{tab:fit}
\end{center}
\end{table*}

\begin{figure}[!htbp]
\centering
\includegraphics[scale=.50]{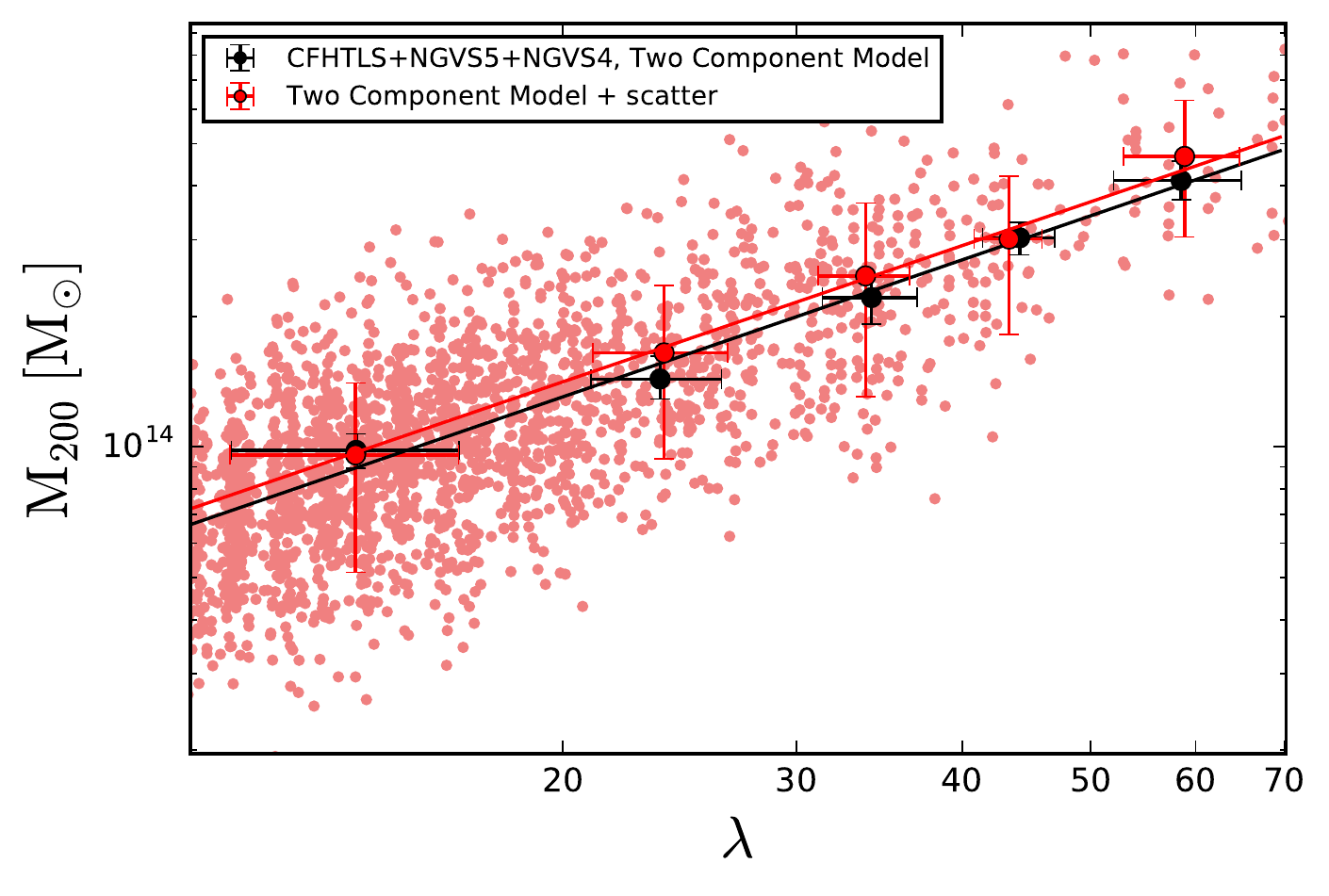}
\caption{\scriptsize Effect of the a posteriori intrinsic scatter correction. Using the mass--richness relation inferred from the Two Component Model (in black), we calculated cluster masses for our selected sample. We scattered those masses, assuming a log-normal distribution centered on $\log{M_{\rm 200}}$ and with a width $\sigma_{\rm \ln{M}|\lambda}=0.39$, based on the value measured by \citet{licitra2016a} (in light red). We repeated this procedure, creating 1000 bootstrap realizations and calculated the new mean mass values in each richness bin, averaging over all realizations. We then repeated the fit to infer the new mass--richness relation (in red), which is shifted toward larger masses.}
\label{fig:scatter}
\end{figure}

				In order to also take into account the intrinsic scatter between richness and mass in the {\it Basic} and in the {\it Two Component Models}, we apply an a posteriori correction as in \citet{ford2015}. Using the mass--richness relation inferred from the {\it Basic Model} and from the {\it Two Component Model}, we calculate the mass of all the clusters in the sample, then we scatter those masses assuming a log-normal distribution centered on $\log{M_{\rm 200}}$ and with a width $\sigma_{\rm \ln{M}|\lambda}=0.39$, based on the scatter measured by \citet{licitra2016a}. We repeat this procedure, creating 1000 bootstrap realizations, choosing masses randomly with replacements from the entire sample. We then calculate the new mean mass values in each richness bin and average them over all bootstrap realizations. We then repeat the fit to infer the new mass--richness relation. This procedure is illustrated in Figure \ref{fig:scatter}, where we show the results from the fit to the {\it Two Component Model} (in black),  the scattered masses (in light red), and the new mean masses and mass--richness relation (in red). Due to the shape of the halo mass function, the net effect of the intrinsic scatter correction is to lead to a slightly higher normalization value of the mass--richness relation. The introduction of the the intrinsic scatter between richness and mass does not significantly change our results obtained with the {\it Basic} or with the {\it Two Component Model}. In fact, the difference in normalization for the original models and their scattered versions is less than $1\%$.
				
				 Having verified the impact of each model term on the final results, we consider as our {\it Final Model} the model that takes into account all the parameters considered so far, the {\it Two Component Model} with the a posteriori intrinsic scatter correction. Our final mass--richness relation is then: $\log{M_{\rm 0}}=14.46\pm0.01$ and $\alpha=1.04\pm0.09$.
				 
				 Our uncertainty on the mass--richness relation parameters above is statistical. We expect systematic biases to be of the same order as the statistical uncertainties, from previous work on the CFHT-LS survey. In fact, \citet{miller2013} and \citet{heymans2012} estimated that the residual bias in the CFHTLenS analysis (and as a consequence on the NGVSLenS, given that the survey characteristics and reduction techniques are the same) could reach maximal values around $3--5\%$ (see also \citet{simet2016}; \citet{fenech2017}), which is on the same order of magnitude of our statistical uncertainties ($\sim$~5\%).

				We checked that our richness binning choice does not affect the recovered mass--richness relation. We performed the fit, discarding the lower (most contaminated) and the highest (less populated) bins, and found consistent results. We have also verified that our procedure does not significantly bias our results, compared to a joint fit \citep[e.g.][]{viola2015, simet2016}. We describe this test in Appendix \ref{appendix}.
												
		\subsection{Comparison with X-Ray Mass Proxies}\label{xray}
		
			To compare our mass estimates with X-ray mass proxies, we follow the same matching procedure as in \citet{licitra2016a}. We use the \citet{gozaliasl2014} and \citet{mehrtens2012} X-ray catalogs, and perform the match between their and our detections imposing a maximum separation of $1~Mpc$ and a maximum difference in redshift of 0.1. We include detections from both the {\it published} and the {\it complete} catalogs to broaden our sample, and have more statistics to perform the scaling relation fits. Results obtained with the {\it complete} catalogs might be affected by contamination biases, since for those, we estimated the purity to decrease to $\sim 60\%$ \citep[Figure 8 and 9 of][]{licitra2016a}.
			\begin{table*}[!htbp]
\begin{center}
\resizebox{.8\textwidth}{!}{
\begin{tabular}{lllllll}
\tableline\tableline\noalign{\smallskip}
Model & $\log{M_{\rm 0}}$ & $\alpha$ & $\left<\rm{diff}\right>_{\rm G}$ & $\left<\rm{diff}\right>_{\rm M}$ & $\left<\rm{M_{\rm L}/M_{\rm X}}\right>_{\rm G}$ & $\left<\rm{M_{\rm L}/M_{\rm X}}\right>_{\rm M}$\\[2pt]
\tableline
\tableline\noalign{\smallskip}
Basic           & $14.43\pm0.01$ & $1.05\pm0.07$ & $0.06\pm0.19$ & $-0.07\pm0.99$ & $1.06\pm0.19$ & $0.93\pm0.99$ \\[5pt]
Basic + ISC & $14.47\pm0.02$ & $1.05\pm0.09$ & $0.17\pm0.20$ & $0.01\pm1.07$ & $1.17\pm0.20$ & $1.01\pm1.07$ \\[5pt]

\tableline\noalign{\smallskip}
Added Scatter           & $14.42\pm0.03$ & $0.97\pm0.14$ & $0.12\pm0.21$ & $-0.07\pm1.06$ & $1.12\pm0.21$ & $0.93\pm1.06$ \\[5pt]

\tableline\noalign{\smallskip}
Two Component           & $14.43\pm0.01$ & $1.05\pm0.07$ & $0.06\pm0.19$ & $-0.07\pm0.99$ & $1.06\pm0.19$ & $0.93\pm0.99$ \\[5pt]
Two Component + ISC & $14.46\pm0.02$ & $1.04\pm0.09$ & $0.15\pm0.20$ & $-0.00\pm1.06$ & $1.15\pm0.20$ & $1.00\pm1.06$ \\[5pt]
\tableline
\end{tabular}
}
\caption{\scriptsize Results of the fit of the mass--richness relation: $\log{M_{\rm 200}}=\log{M_{\rm 0}}+\alpha\log{\lambda/\lambda_{\rm 0}}$, with a pivot $\lambda_{\rm 0}=40$, obtained using the three models. For the Basic and Two Component Models, we also show the results after applying the a posteriori intrinsic scatter correction (ISC). The last four columns show the normalized average difference between lensing and X-ray masses, $<\rm{diff}>=\left(M_{\rm L}-M_{\rm X}\right)/M_{\rm X}$, and the average ratio of the two, $<\rm{M_{\rm L}/M_{\rm X}}>$, using the X-ray detections of \citet{gozaliasl2014} (G), and \citet{mehrtens2012} (M).}
\label{tab:mr_par}
\end{center}
\end{table*}

			Within all three fields, we recover 36(27) objects from the match of the {\it complete}({\it published}) catalog with  \citet{gozaliasl2014}  (in this case, all objects are from the CFHT-LS W1 field), and 21(17) from objects from the match of the {\it complete}({\it published}) catalog with  \citet{mehrtens2012}. As shown in \citet{licitra2016a}, RedGOLD recovers 38 clusters, up to $z\sim1$, in the $3~deg^2$ of the CFHT-LS W1 field, covered by \citet{gozaliasl2014} catalog. The clusters detected by RedGOLD that do not have an X-ray counterpart seem to be, from visual inspection, small galaxy groups. It is possible that these systems have an X-ray emission below the X-ray detection limit, or that they are not relaxed systems and do not have any X-ray emission at all.
			
			As explained in Section \ref{xray_cat}, \citet{gozaliasl2014} $M_{\rm 200}$ masses were estimated using the $M_{\rm X}-L$ relation of \citet{leauthaud2010}. We estimate \citet{mehrtens2012} $M_{\rm 200}$ masses from the $r_{\rm 200}$ values given in their catalog, using Equation \ref{mass}. Our masses $M_{\rm 200}^{lens}$ are calculated using our final mass--richness relation.
			
\begin{figure*}[!htbp]
\centering
\includegraphics[scale=.50]{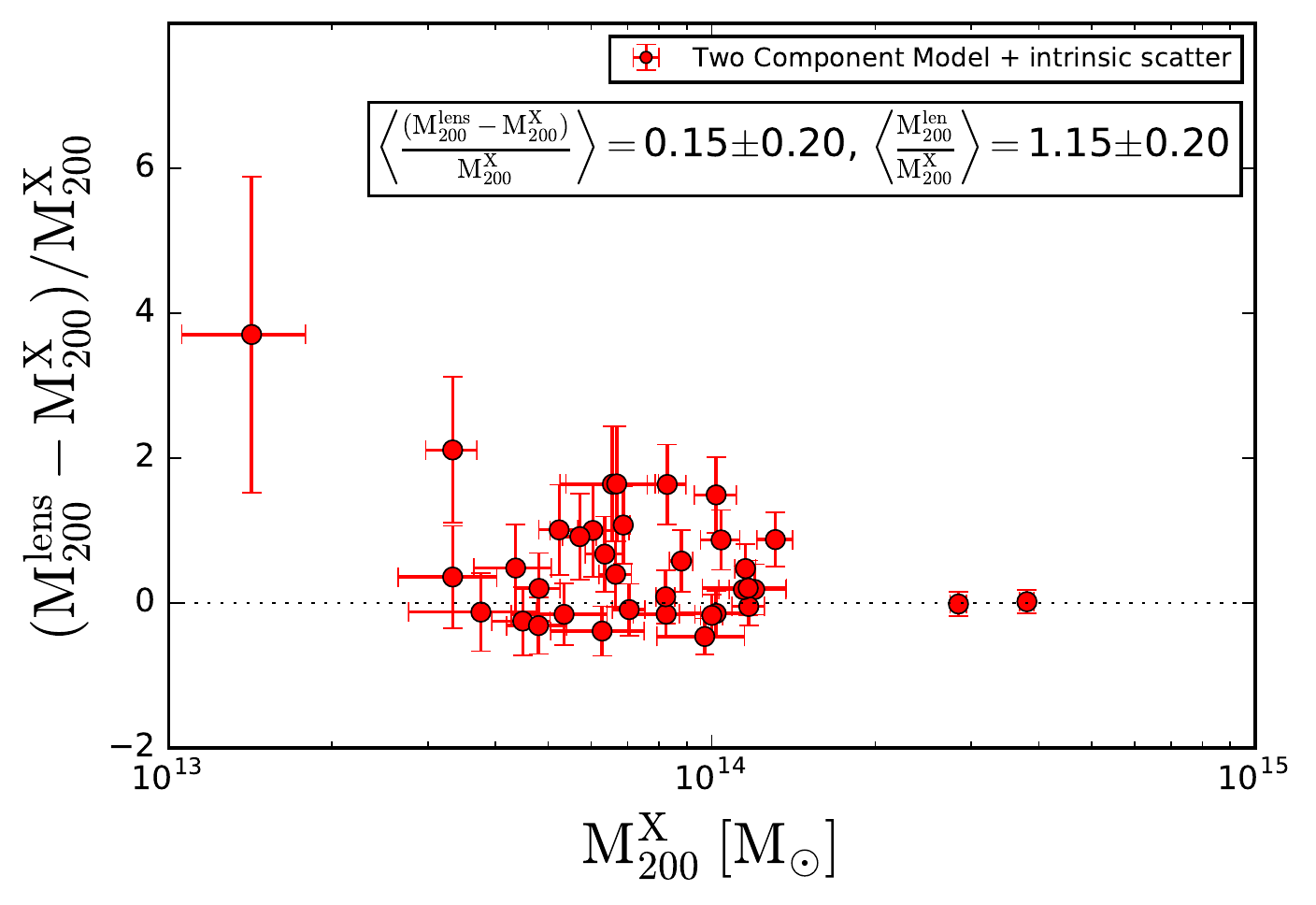}
\includegraphics[scale=.50]{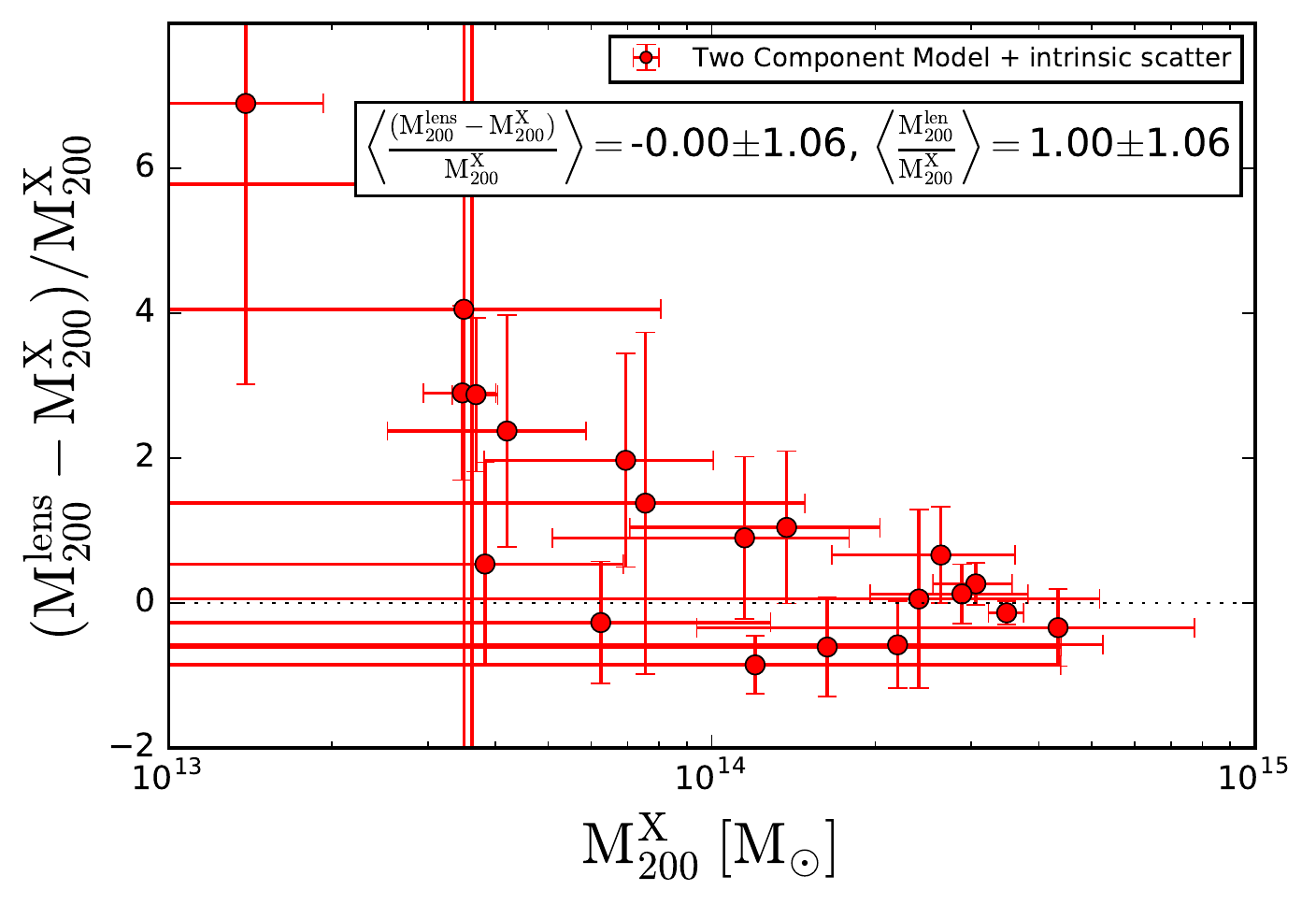}
\caption{\scriptsize Comparison of lensing masses and X-ray masses calculated from the fitted mass--richness relations obtained using our {\it Final Model} (i.e. the Two Component Model with the a posteriori intrinsic scatter correction). Using \citet{gozaliasl2014} catalog (on the left), we obtain a mean normalized difference of $0.15\pm0.20$ and a mean ratio of $1.15\pm0.20$, while using \citet{mehrtens2012} catalog (on the right), we find $-0.00\pm1.06$ and $1.00\pm1.06$, respectively.}
\label{fig:xray_res}
\end{figure*}

			In Figure \ref{fig:xray_res}, we show the normalized difference between the X-ray masses of \citet{gozaliasl2014} and lensing masses $\left(M_{\rm 200}^{lens}-M_{\rm 200}^{X}\right)/M_{\rm 200}^{X}$ as a function of $M_{\rm 200}^{X}$, obtained using our {\it Final Model}.  The ratio is measured with respect to $M_{\rm 200}^{X}$ since our sample is X-ray selected (i.e. we select the clusters in the X-ray catalog, and then compare their X-ray and lensing mass estimate).
			
			In the last four columns of Table \ref{tab:mr_par}, we show the mean normalized difference and the mean ratio between lensing and X-ray masses, for the three models, obtained with \citet{gozaliasl2014} and \citet{mehrtens2012} catalogs. For all models, the mean differences obtained using $M_{\rm 200}^{X}$ from \citet{gozaliasl2014} ($\sim0.1-0.2$) are higher than those obtained using \citet{mehrtens2012} ($\sim -0.1-0.0$). However, uncertainties on the individual measurements are larger and the scatter in the difference are about an order of magnitude higher for the \citet{mehrtens2012} sample. Because of the large uncertainty, we do not consider results obtained with the \citet{mehrtens2012} catalogs reliable.
			
			As explained in Section \ref{xray_cat}, \citet{gozaliasl2014} masses were calculated from the X-ray luminosity, after the excision of the AGN contribution and the correction for cool core flux removal. We find that this leads to X-ray mass estimates that are lower compared to masses derived with weak lensing than those calculated without core excision.  Hereafter, we will use only the \citet{gozaliasl2014} sample, given the larger uncertainty in our results obtained using the  \citet{mehrtens2012} sample, and the higher number of cluster matches. Core-excised X-ray temperatures are also known to better correlate with cluster masses \citep{pratt2009}.
			
			 Using X-ray masses from the \citet{gozaliasl2014} catalog and the lensing masses estimated from the mass--richness relation derived from our {\it Final Model}, applied on the {\it complete} catalogs, we find a mean normalized difference of  $0.15\pm0.20$ ($\frac{M_{\rm 200}^{lens}}{M_{\rm 200}^{X}}=1.15\pm0.20$), considering the whole mass range. If we consider two different mass ranges, we find a mean normalized difference of $0.17\pm0.24$ for $M_{\rm 200}^{X}<10^{14}M_{\rm \odot}$, and a mean normalized difference of $0.14\pm0.18$ for $M_{\rm 200}^{X}\geq10^{14}M_{\rm \odot}$. This corresponds to $\sim15\%$ higher lensing masses in the whole mass range, and $\sim20\%$ and $\sim15\%$ higher lensing masses for $M_{\rm 200}^{X}<10^{14}M_{\rm \odot}$ and $M_{\rm 200}^{X}>10^{14}M_{\rm \odot}$, respectively.
			 		
			To obtain scaling relations, we exclude the two clusters with mass $M_{\rm 200}^{X}< 2 \times10^{13}M_{\rm \odot}$ from the matched sample with \citet{gozaliasl2014}, because both our and the X-ray catalog are incomplete at these low masses. We also do not consider the two highest mass matches ($M_{\rm 200}^{X}>2 \times10^{14}M_{\rm \odot}$), because our catalog is incomplete in this mass range, given our low area coverage. All four excluded clusters were matches with the Licitra's {\it published} catalog.
			
					\begin{figure}[!htbp]
\centering
\includegraphics[scale=.50]{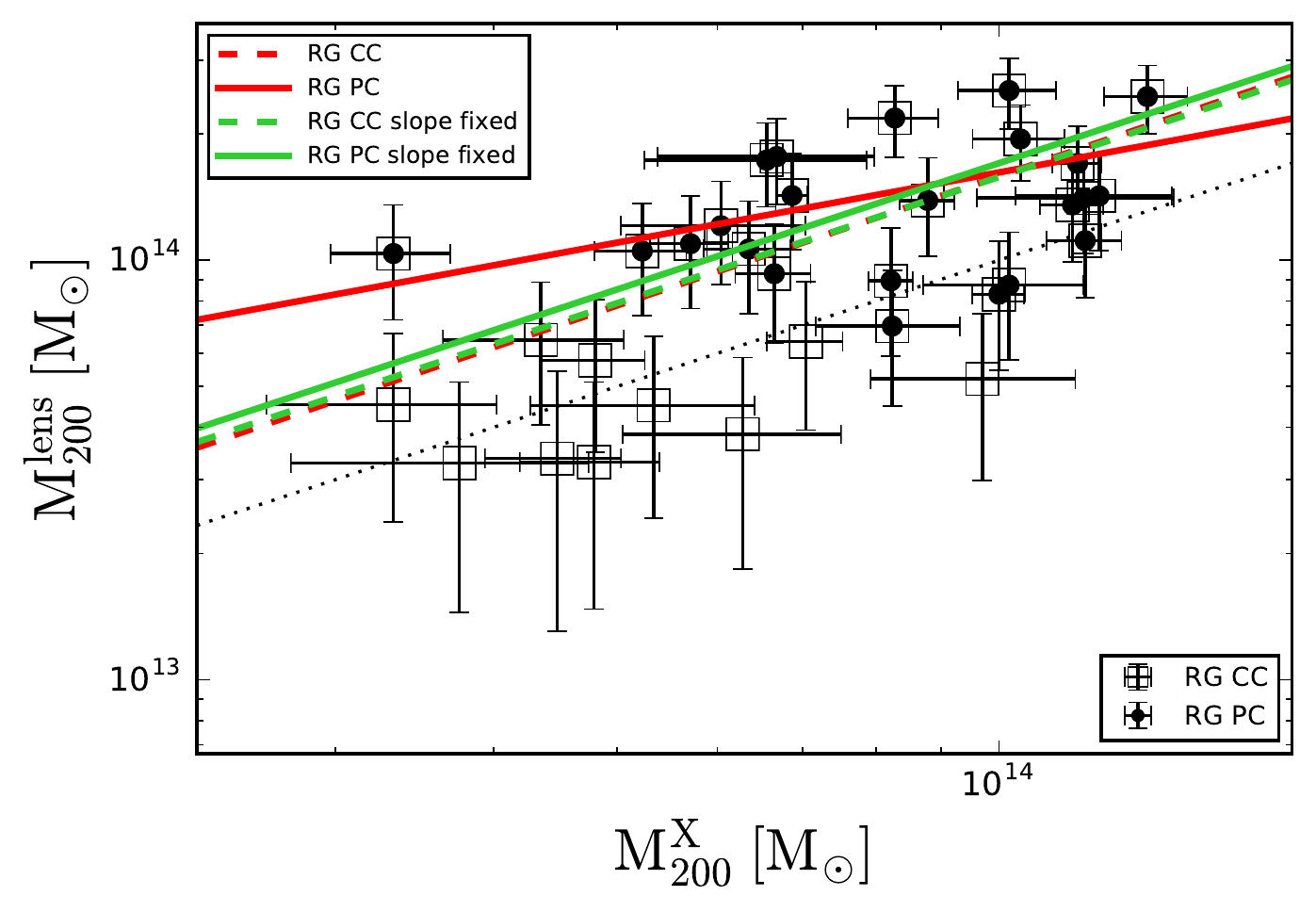}
\caption{\scriptsize We compare our derived weak lensing masses with the X-ray masses from  \citet{gozaliasl2014} catalog. The weak lensing masses have been derived from our fit of the mass--richness relation using our {\it Final Model}. The black dots are the RedGOLD detections from the {\it published} catalogs (RG PC) and the black squares are the detections from the {\it complete} catalogs (RG CC). The red lines show the fits obtained with the slope as a free parameter, and the green lines those obtained with the slope fixed at unity. In both cases, solid lines refer to the {\it published} catalogs, and the dashed lines to the {\it complete} catalogs. The black dotted line is the diagonal. See Section \ref{redgold_cat} for the catalog definitions.}
\label{fig:xray_scal1}
\end{figure}

\begin{figure*}[!htbp]
\centering
\includegraphics[scale=.50]{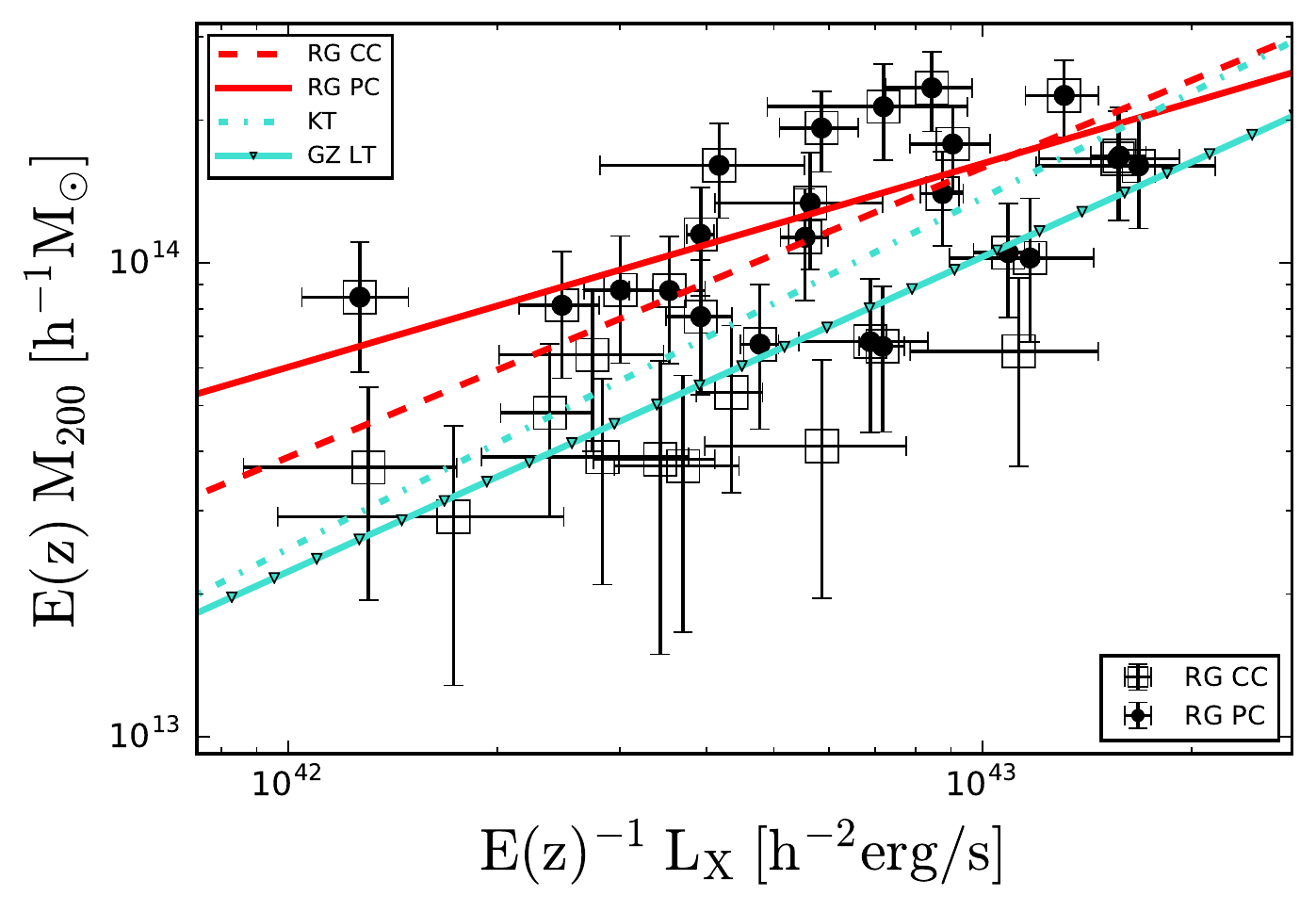}
\includegraphics[scale=.50]{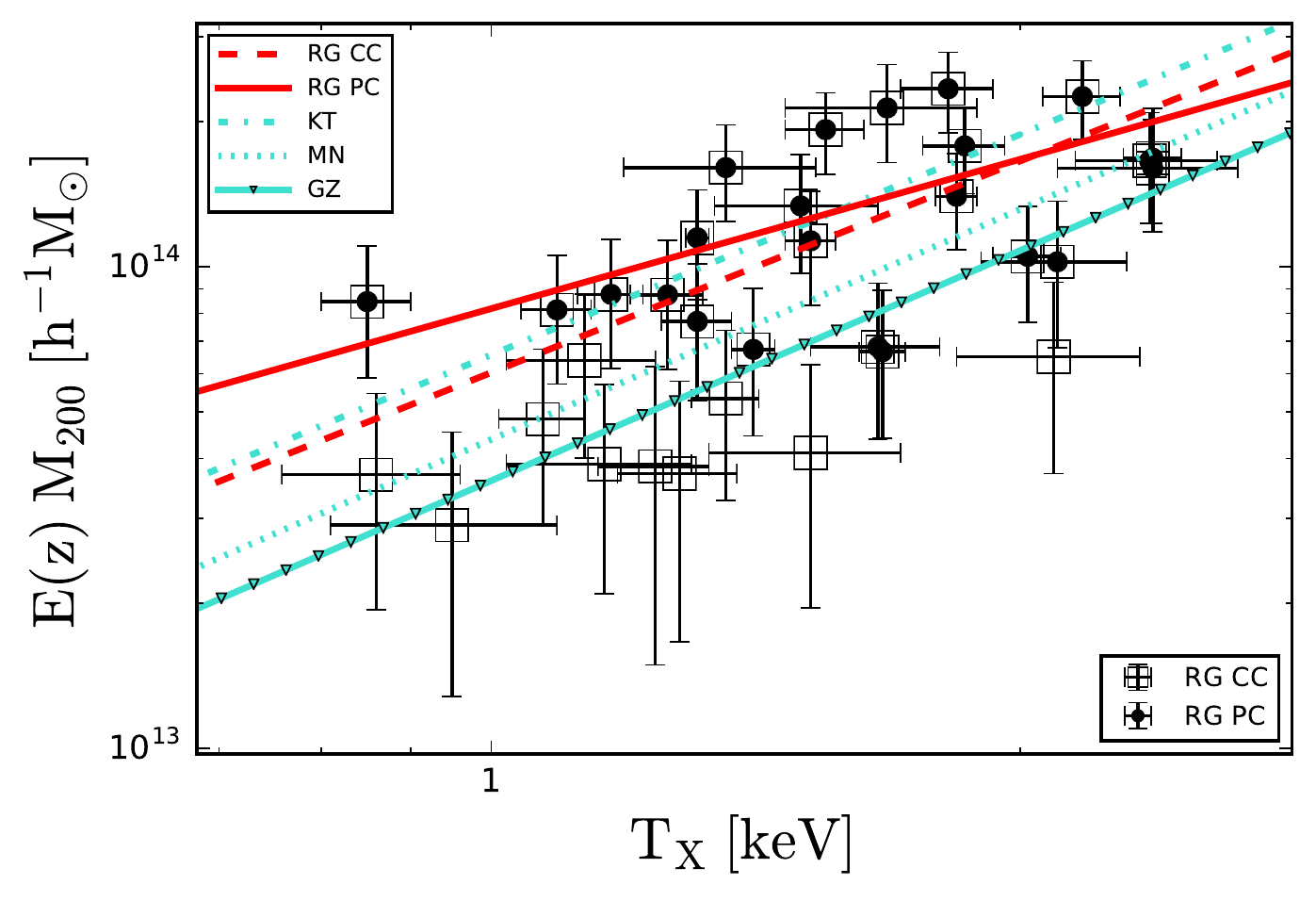}
\caption{\scriptsize We compare our derived weak lensing masses with X-ray mass proxies. On the left, we show our mass-luminosity relation, and on the right, our mass--temperature relation, compared with the literature. The black dots and the black squares have the same meaning as in Figure \ref{fig:xray_scal1}. The different cyan lines show the relations obtained using the X-ray masses from \citet{gozaliasl2014} catalog (GZ), calculated with the M--L relation of \citet{leauthaud2010} (LT), and the relations inferred by \citet{kettula2015} (KT) and by \citet{mantz2016} (MN) (see the legend for the line styles).}
\label{fig:xray_scal2}
\end{figure*}

			In Figure \ref{fig:xray_scal1}, we plot the  $M_{\rm 200}^{X}-M_{\rm 200}^{lens}$ relation, and in Figure \ref{fig:xray_scal2}, the $L_{\rm X}-M_{\rm 200}^{lens}$ and the $T_{\rm X}-M_{\rm 200}^{lens}$ relations. In those plots, the black dots represent matches with the RedGOLD cluster detections in Licitra's  {\it published} catalogs, while the black squares represent all those with the  {\it complete} catalogs (see Section \ref{redgold}).  This difference between our lensing masses and those calibrated with lensing masses from \citet{leauthaud2010} includes different contributions, and it is not a straightforward difference between our lensing masses and X-ray selected lensing masses. In fact, both the \citet{gozaliasl2014} selection in $L_X$ (when stacking clusters to derive the \citet{leauthaud2010} lensing masses), our selection based on the \citet{licitra2016a,licitra2016b} richness, and differences in the shear calibration in our data and  \citet{leauthaud2010} contribute to this difference, and interpreting it precisely implies degeneracies on each contribution.

			In Figure \ref{fig:xray_scal1}, we show the relation between X-ray and lensing masses:
		
			\begin{subequations}
				\begin{align}
					\log{\left({M_{\rm 200}^{lens}}\right)} & =a+b \log{\left({M_{\rm 200}^{X}}\right)}
				\end{align}
			\end{subequations}
			
			The black dotted line is the diagonal, the solid lines are the fit to the {\it published} catalogs, and the dashed lines are the fit to the {\it complete} catalogs. The red lines were obtained with the slope as a free parameter of the fit, and the green lines with the slope fixed at unity. For the {\it published } catalogs, our threshold in richness and $\sigma_{\rm det}$ is meant to select clusters with $M_{\rm 200}>10^{14}M_{\rm \odot}$ with a completeness $\sim80\%$. Part of these detections have X-ray masses lower than our selection threshold of $M_{\rm 200}>10^{14}M_{\rm \odot}$; in fact, their X-ray masses are in the range $ 2 \times 10^{13}M_{\rm \odot}< M_{\rm 200}^{X}<10^{14}M_{\rm \odot}$. We expect to have a contamination of clusters with these lower masses, and our purity of $\sim 80\%$ is calculated for real clusters with $M_{\rm 200}^{X}>10^{13}M_{\rm \odot}$. However, our completeness decreases (<80\%) in this mass range ($M_{\rm 200}^{X}<10^{14}M_{\rm \odot}$), as shown in \citet{licitra2016a}. 
			
			\begin{table}[!htbp]
\begin{center}
\resizebox{.48\textwidth}{!}{
\begin{tabular}{llrrl}
\tableline\tableline\noalign{\smallskip}
Relation & Sample & a & b & scatter\\[2pt]
\tableline
\tableline\noalign{\smallskip}
\multirow{2}{*}{$M_{\rm L}-M_{\rm X}$} & CC & $-0.13\pm2.96$ & $1.02\pm0.21$ & 0.20\\[5pt]
						& PC & $6.42\pm3.17$ & $0.56\pm0.23$ & 0.15\\[5pt]
\hline
\multirow{2}{*}{$M_{\rm L}-M_{\rm X}$} & CC & $0.20\pm0.03$ & fixed at 1 & 0.20\\[5pt]
						& PC & $0.23\pm0.03$ & fixed at 1 & 0.17\\[5pt]
\hline
\multirow{2}{*}{$M_{\rm L}-T_{\rm X}$} & CC & $0.23\pm0.03$ & $1.46\pm0.28$ & 0.20 \\[5pt]
						& PC & $0.28\pm0.03$ & $1.03\pm0.30$ & 0.15\\[5pt]
\hline
\multirow{2}{*}{$M_{\rm L}-L_{\rm X}$} & CC & $0.10\pm0.03$ & $0.61\pm0.12$ & 0.20\\[5pt]
						& PC & $0.16\pm0.03$ & $0.43\pm0.12$ & 0.15\\[5pt]
\tableline
\end{tabular}
}
\caption{\scriptsize Results of the fit of the weak lensing mass vs X-ray mass and mass proxy relations: $\log{M_{\rm L}=a+b\log{M_{\rm X}}}$; $\log{\left(M_{\rm 200} E(z)/M_{\rm 0}\right)}=a+b \log{\left(L_{\rm X}/L_{\rm 0} E(z)\right)}$; $\log{\left(M_{\rm 200}E(z)/M_{\rm 0}\right)}=a+b \log{\left(T_{\rm X}/T_{\rm 0}\right)}$. "CC" refers to the {\it complete catalogs} and "PC" to the {\it published catalogs} (see text for the catalogs definitions and for the values of the pivot mass, luminosity, and temperature used in the fit of the scaling relations).}
\label{tab:xray}
\end{center}
\end{table}

			 When fixing the slope at the unity, we obtain $a=0.20\pm0.03$($a=0.23\pm0.03$), and a scatter of $\sigma_M = 0.20$~dex ($\sigma_M = 0.17$~dex) for the {\it complete} ({\it published}) catalogs. In this case, the difference in $a$ for the two samples is negligible, $\sim 0.03 \pm0.06$~dex. The small shift in normalization ($\sim 0.2$~dex) compared to the diagonal is expected, because lensing mass estimates are generally higher than X-ray masses \citep{zhang2008, rasia2012, simet2015}. When leaving the slope as a free parameter, we find $a=-0.13\pm2.96$ and $b=1.02\pm0.21$, with a scatter of $\sigma_M = 0.20$~dex ($a=6.42\pm3.17$ and $b=0.56\pm0.23$, with a scatter of $\sigma_M = 0.15$~dex) for the {\it complete} ({\it published}) catalogs. The incompleteness when using the {\it published} catalogs appears to bias our fit slope, which becomes much shallower than the diagonal.	
		
			In Figure \ref{fig:xray_scal2}, we show the mass--luminosity and mass--temperature relations. We apply a logarithmic linear fit, in the form:\\
			\begin{subequations}
				\begin{align}
					\log{\left(\frac{M_{\rm 200} E(z)}{M_{\rm 0}}\right)} & =a+b \log{\left(\frac{L_{\rm X}}{L_{\rm 0} E(z)}\right)}\\
					\log{\left(\frac{M_{\rm 200} E(z)}{M_{\rm 0}}\right)} & =a+b \log{\left(\frac{T_{\rm X}}{T_{\rm 0}}\right)}
				\end{align}
			\end{subequations}
			where $E(z)=H(z)/H_{\rm 0}$, $M_{\rm 0}=8\times10^{13}\;h^{-1}M_{\rm \odot}$ for the $M_{\rm 200}-L_{\rm X}$, $M_{\rm 0}=6\times10^{13}\;h^{-1}M_{\rm \odot}$ for the $M_{\rm 200}-T_{\rm X}$, $L_{\rm 0}=5.6\times10^{42}\;h^{-2}erg/s$, and $T_{\rm 0}=1.5\;keV$.

			For the mass--luminosity relation, we find $a=0.10\pm0.03$ and $b=0.61\pm0.12$, with a scatter $\sigma_{\rm logM_{\rm 200}|L_X}=0.20$~dex($a=0.16\pm0.03$ and $b= 0.43\pm0.12$, with a scatter $\sigma_{\rm logM_{\rm 200}|L_X}=0.15$~dex) for the {\it complete}({\it published}) catalogs. For the mass--temperature relation, we find $a=0.23\pm0.03$ and $b=1.46\pm0.28$, with a scatter $\sigma_{\rm logM_{\rm 200}|T_X}=0.20$~dex($a=0.28\pm0.03$ and $b=1.03\pm0.30$, with a scatter $\sigma_{\rm logM_{\rm 200}|T_X}=0.15$~dex), for the {\it complete}({\it published} catalogs). The relations obtained with the {\it published} catalogs show again shallower slopes. Our results are consistent with the expected deviations from self-similarity \citep{bohringer2011}.
				
			We summarize our results in Table~\ref{tab:xray}.

\section{DISCUSSION}\label{disc}

	\subsection{Comparison to Previously Derived Mass--Richness Relations}
		
		In this section, we discuss our results in the context of similar current studies.
		
		As stated before and shown in \citet{licitra2016a, licitra2016b}, our richness estimator $\lambda$ is defined in a similar way as the richness from redMaPPer  \citep{rykoff2014}. The redMaPPer richness is defined as $\lambda_{\rm RM}=\sum p_{\rm mem}\theta_{\rm L}\theta_{\rm R}$, where $p_{\rm mem}$ is the probability that each galaxy in the vicinity of the cluster is a red-sequence member and $\theta_{\rm L}, \theta_{\rm R}$ are weights that depend on luminosity and radius. In this calculation, only galaxies brighter than $0.2L_{\rm *}$ and within a scale radius $R_{\rm \lambda}$ are considered. The radius is richness-dependent and it scales as $R_{\rm \lambda}=1.0(\lambda/100)^{0.2}h^{-1}Mpc$.
						    
		The RedGOLD richness is a simplified version of $\lambda_{\rm RM}$. We constrained the radial distribution of the red-sequence galaxies with an NFW profile and applied the same luminosity cut and radius scaling as in \citet{rykoff2014} but did not apply a luminosity filter. Unlike the redMaPPer definition, our richness is not a sum of probabilities. Those choices were made to minimize the scatter in the mass--richness relation. For redshifts $z<0.3$, the difference $\frac{\lambda_{\rm RM} - \lambda}{\lambda}$ is only of $5--15\%$, while it increases to $40--60\%$ at $0.4<z<0.5$, where the redMaPPer richness is systematically higher \citep{licitra2016a}. This difference might be due to the different depths of the CFHTLenS and SDSS surveys. This means that we can compare our results with others obtained using the redMaPPer cluster sample. 

		\citet{simet2016} performed a stacking analysis of the redMaPPer cluster sample, using shear measurements from the SDSS. Their sample is much larger than ours, consisting of 5,570 clusters, with a redshift range $0.1<z<0.3$, lower than the one used for this work,  and a richness range $20\leq \lambda_{\rm RM} \leq 140$. With these data, they were able to characterize the different systematic errors arising in their analysis with great accuracy. For the mass--richness relation, they obtained the normalization $\log{(M_{\rm 0}\;[h^{-1}M_{\rm \odot}])}=14.34\pm0.04$ (the error includes both statistical and systematic error) and the slope $\alpha=1.33^{+0.9}_{\rm -0.1}$. To compare our results to theirs, we use our masses in units of $h^{-1}M_{\rm \odot}$ and we repeat our fits. Using our {\it Final Model}, we obtain $\log{M_{\rm 0}}=14.31\pm0.02$ and $\alpha=1.04\pm0.09$ (the errors are only statistical). Our normalization is consistent within $1 \sigma$ and our slope is consistent within $\sim2\sigma$ of Simet et al.'s. Comparing the masses at the pivot richness, $\lambda_{\rm 0}=40$, we obtain $2.04\times10^{14}h^{-1}M_{\rm \odot}\pm0.02$ compared to Simet et al.'s  $2.21\times10^{14}h^{-1}M_{\rm \odot}\pm0.15$.
	
		In another recent work, \citet{farahi2016} inferred the mass--richness relation using the same sample of SDSS redMaPPer clusters ($0.1<z<0.3$ and $\lambda_{\rm RM}>20$), performing a stacking analysis and estimating the velocity dispersion of the dark matter halos from satellite-central galaxy pairs measurements. For the mass--richness relation, they found a normalization of $14.19\pm0.1$ and a slope of $1.31\pm0.19$ (the error includes both statistical and systematic error), using a pivot $\lambda_{\rm 0}=30$. Repeating the fit using their pivot richness, we obtain $\log{M_{\rm 0}}=14.18\pm0.02$ and $\alpha=1.04\pm0.09$, consistent within less than $1 \sigma$ in normalization and $1.5\sigma$ in slope, with their results. At the pivot richness $\lambda_{\rm 0}=30$ our mass is $1.51\times10^{14}M_{\rm \odot}\pm0.02$, consistent with their value of $1.56\times10^{14}M_{\rm \odot}\pm0.35$.

		\citet{melchior2016} calibrated the mass--richness relation and its evolution with redshift up to $z<0.8$, using 8000 RedMaPPer clusters in the Dark Energy Survey Science Verification \citep[DES;][]{des2016} with $5\leq\lambda_{\rm RM}\leq180$. They found a normalization $M_{\rm 0}=2.35\pm0.34\times10^{14}M_{\rm \odot}$ and a slope $1.12\pm0.26$, using the pivot richness $\lambda_{\rm 0}=30$ and a mean redshift $z=0.5$. Their errors include both statistical and systematic errors. These results are consistent with ours within less than $1 \sigma$, both in normalization and slope, even if this sample has a larger average redshift, where we expect our richness definitions to be less similar. 
		
		\begin{table*}[!htbp]
\begin{center}
\resizebox{.8\textwidth}{!}{
\begin{tabular}{l|lccccc}
\tableline\tableline\noalign{\smallskip}
Relation & Comparison & Sample & $\Delta a$ & $\Delta b$ & a Compatibility & b Compatibility\\[2pt]
\tableline
\tableline\noalign{\smallskip}
\multirow{4}{*}{$M_{\rm L}-T_{\rm X}$} & \multirow{2}{*}{\citet{kettula2015}} & CC & $0.08\pm0.14$ & $0.06\pm0.33$ & $1~\sigma$ & $1~\sigma$\\[5pt]
							     &	                                         		& PC & $0.25\pm0.15$ & $0.49\pm0.34$ & $2~\sigma$ & $1.5~\sigma$\\[5pt]
\cline{2-7}								
							     & \multirow{2}{*}{\citet{mantz2016}} & CC & $0.27\pm0.28$ & $0.06\pm0.13$ & $1~\sigma$ & $1~\sigma$\\[5pt]
							     &							   & PC & $0.22\pm0.54$ & $0.34\pm0.28$ & $1~\sigma$ & $1.5~\sigma$\\[5pt]

\hline
\multirow{4}{*}{$M_{\rm L}-L_{\rm X}$} & \multirow{2}{*}{\citet{kettula2015}} & CC & $0.06\pm0.15$ & $0.13\pm0.15$ & $1~\sigma$ & $1~\sigma$\\[5pt]
							     &	                                                   & PC & $0.24\pm0.15$ & $0.31\pm0.15$ & $1.5~\sigma$ & $2~\sigma$\\[5pt]
\cline{2-7}								
							     & \multirow{2}{*}{\citet{leauthaud2010}} & CC & $0.22\pm0.08$ & $0.05\pm0.18$ & $2.5~\sigma$ & $1~\sigma$\\[5pt]
							    &								     & PC & $0.33\pm0.09$ & $0.23\pm0.19$ & $4~\sigma$ & $1.5~\sigma$\\[5pt]
\tableline
\end{tabular}
}
\caption{\scriptsize Comparison of our mass--temperature and mass--luminosity relations with others in literature.  "CC" refers to the results obtained using the {\it complete catalogs} and "PC" using the {\it published catalogs} (see text for the catalogs definitions). Here, $\Delta a$ is the difference in normalization, and $\Delta b$ the difference in slope, between our results and those obtained by \citet{kettula2015}, \citet{mantz2016} and \citet{leauthaud2010}. The last two columns show that our relations are consistent, in normalization and slope, within $\lesssim 1~\sigma$ with the others in literature ($\lesssim 2.5~\sigma$ in normalization with \citet{leauthaud2010}), when using the complete catalogs.}
\label{tab:xray2}
\end{center}
\end{table*}

		Our normalization is in perfect agreement with all the works cited above ($<1\sigma$). On the other hand, there is a slight tension between our slope and those of \citet{simet2016} and \citet{farahi2016} ($1.5-2\sigma$), but not with \citet{melchior2016} ($<1\sigma$). Our slope is also consistent with the first mass--richness relation inferred using the redMaPPer cluster sample from \citet{rykoff2012}, and with the \citet{saro2015} richness-mass relation, inferred by cross-matching the SPT-SZ survey with the DES redMaPPer cluster sample. They found values of 1.08 (the error is not given) and $0.91\pm0.18$, respectively. \citet{saro2015} value has been converted from the slope of the richness-mass relation to the slope of the mass--richness relation by \citet{simet2016}.
		
		We cannot compare our results with the scaling relations obtained in \citet{johnston2007}, \citet{covone2014}, \citet{ford2015}, or \citet{vanuitert2015} because their definition of richness is different.
			
		We conclude that our fit of the mass--richness relation is in agreement with the other works cited above. These results confirm the efficiency of the RedGOLD richness estimator, and quantify the relation between the RedGOLD richness measurements and the total cluster masses obtained with weak lensing. Even without using a probability distribution, our richness is as efficient as the more sophisticated redMaPPer richness definition.

	\subsection{Weak Lensing vs X-Ray Masses}\label{xray2}

		In Figure \ref{fig:xray_scal2}, we compare our lensing mass versus X-ray mass proxies relations to those of other works in literature. 
	
		In the $L_{\rm X}-M_{\rm 200}^{lens}$ plot, we compare our results with those from \citet{kettula2015} and \citet{leauthaud2010}. We remind the reader that the fit to the {\it published} catalogs (solid red line) shows a shallower slope because of our selection in mass, which, while it optimizes purity, leads to a bias in slope due to the lack of clusters detected at masses $M_{\rm 200}<10^{14}M_{\rm \odot}$ (see discussion in Section~\ref{xray}).
	
		Because of the large uncertainties, the fit to both the {\it complete} and  {\it published} catalogs (dashed red line) are consistent within $<1 \sigma$ and $<2 \sigma$, respectively, in normalization and slope, with results from \citet{kettula2015}, even if our normalizations are higher. 
					
		With respect to the $E(z) M_{\rm 200}$ derived from \citet{leauthaud2010} (and, as a consequence, from  \citet{gozaliasl2014}, because they use \citet{leauthaud2010} to derive their mass relations), we are consistent within $<2.5 \sigma$ in normalization and within $<1 \sigma$ in slope for the {\it complete} catalogs. For the {\it published} catalogs, we are inconsistent in normalization (the normalization difference is $\sim 3.7\sigma$) but consistent in slope within $<1.5 \sigma$.

		Both \citet{kettula2015} and \citet{leauthaud2010} did not apply the miscentering correction but, while the first performed their lensing analysis on single clusters, the latter stacked their low-mass clusters in very poorly populated bins. This procedure could have introduced a bias that led to more smoothed profiles and thus to lower mass estimates and to a lower normalization of the scaling relation.
						
		 In the $T_{\rm X}-M_{\rm 200}^{lens}$ plot, we compare our results with \citet{kettula2015} and \citet{mantz2016}. Because their masses are derived at the overdensity $\Delta=500$, we convert their $M_{\rm 500}$ values to $M_{\rm 200}$, using $M_{\rm 200}=1.35M_{\rm 500}$ from \citet{rines2016}, derived considering that the mass--concentration relation weakly depends on mass \citep{bullock2001} and assuming an NFW profile with a fixed concentration $c=5$. We find that the normalization and slope of our fit to the {\it complete}({\it published}) catalogs are consistent with the \citet{kettula2015} results within $<1\sigma$($ \lesssim2\sigma$), and with \citet{mantz2016} results within $<1\sigma$($<1.5\sigma$) in normalization and slope. 
		 
		 In Table \ref{tab:xray2}, we show the differences in normalization, $\Delta a$, and in slope, $\Delta b$, between our results and those used for comparison for the mass--luminosity, and the mass--temperature relations.
		 				 			 
	 	Given that our results based on the RedGOLD {\it complete} catalogs are consistent with other results in the literature, we conclude that the thresholds that we apply in the RedGOLD {\it published} catalog introduces systematics in the fit of the cluster lower mass end. 

		Selecting samples based on lensing measurements, simulations predict that mass measurements from lensing are systematically lower than the cluster true total mass by $\sim 5--10\%$ (in the mass range $M^{sim}>5\times 10^{14}$) and those from X-ray proxies (in the mass range $10^{14}<M_{\rm 200}^{X}<5\times 10^{15}$) by $\sim 25--35\%$, with $\left<M^{sim}_{\rm {X}}/M^{sim}_{\rm {L}}\right> \sim 0.7-0.8$  \citep{meneghetti2010, rasia2012}. When we compare our weak lensing mass measurements to X-ray \citet{gozaliasl2014} cluster masses (Figure~\ref{fig:xray_res} and  Table \ref{tab:mr_par}), for X-ray selected clusters, for the {\it Final Model} we obtain $\sim15\%$ higher lensing masses in the whole mass range, and $\sim20$ and $\sim15\%$ higher lensing masses for $M_{\rm 200}^{X}<10^{14}M_{\rm \odot}$ and $M_{\rm 200}^{X}>10^{14}M_{\rm \odot}$, respectively. 
				
		As we mentioned before in Section \ref{xray}, and from Table \ref{tab:mr_par} and Figure \ref{fig:xray_res}, the mean residuals and ratio values obtained using \citet{mehrtens2012} catalog are lower, with $\left<M_{\rm {L}}/M_{\rm {X}}\right> \sim -0.1-0.0$, which means that non core-excised temperature led to overestimated X-ray masses, as expected \citep{pratt2009}.
		
		\begin{table*}[!htbp]
		\begin{center}
		\resizebox{.6\textwidth}{!}{
			\begin{tabular}{llllllll}
				\hline\hline\noalign{\smallskip}
				Model & $\log M_{0}$ & $\alpha$ & $p_{cc}$ & $\sigma_{off}$ & $\sigma_{M|\lambda}$ & aMbcg & aCM\\[2pt]
				\hline
				\hline\noalign{\smallskip}
				1 & (11,16) & (-2, 2) & (0, 1) & (0, 2) & -- & -- & --\\[5pt]
				2 & (11,16) & (-2, 2) & (0, 1) & (0, 2) & (0.1, 0.7) & -- & --\\[5pt]
				3 & (11,16) & (-2, 2) & (0, 1) & (0, 2) & -- & (0, 10) & --\\[5pt]
				4 & (11,16) & (-2, 2) & (0, 1) & (0, 2) & -- & -- & (0, 10)\\[5pt]
				5 & (11,16) & (-2, 2) & (0, 1) & (0, 2) & (0.1, 0.7) & (0, 10) & (0, 10)\\[5pt]
				\hline
			\end{tabular}
		}
		\caption{\scriptsize MCMC uniform prior ranges for the different parameters of the five models of the joint fit, described in Appendix \ref{appendix}. The lack of a numerical value indicates that the parameter is not included in the respective model.}
	\label{tab:priors_joint}
	\end{center}
\end{table*}

\begin{table*}[!htbp]
	
		\begin{center}
		\resizebox{.7\textwidth}{!}{
			\begin{tabular}{llllllll}
				\hline\hline\noalign{\smallskip}
				Model & $\log M_{0}$ & $\alpha$ & $p_{cc}$ & $\sigma_{off}$ & $\sigma_{M|\lambda}$ & aMbcg & aCM\\[2pt]
				\hline
				\hline\noalign{\smallskip}
				1 & $14.49^{+0.03}_{-0.03}$ & $1.28^{+0.06}_{-0.06}$ & $0.65^{+0.05}_{-0.05}$ & $0.9^{+0.5}_{-0.3}$ & -- & -- & --\\[5pt]
				2 & $14.49^{+0.05}_{-0.05}$ & $1.27^{+0.15}_{-0.13}$ & $0.65^{+0.04}_{-0.06}$ & $1.0^{+0.5}_{-0.4}$ & $0.12^{+0.05}_{-0.01}$ & -- & --\\[5pt]
				3 & $14.48^{+0.03}_{-0.03}$ & $1.28^{+0.06}_{-0.06}$ & $0.65^{+0.05}_{-0.06}$ & $0.8^{+0.5}_{-0.4}$ & -- & $4^{+4}_{-3}$ & --\\[5pt]
				4 & $14.41^{+0.02}_{-0.02}$ & $1.19^{+0.06}_{-0.06}$ & $0.83^{+0.10}_{-0.18}$ & $0.6^{+0.8}_{-0.4}$ & -- & -- & $0.65^{+0.07}_{-0.06}$\\[5pt]
				5 & $14.40^{+0.04}_{-0.06}$ & $1.17^{+0.14}_{-0.12}$ & $0.83^{+0.11}_{-0.16}$ & $0.6^{+0.7}_{-0.4}$ & $0.13^{+0.07}_{-0.02}$ & $2^{+4}_{-1}$ & $0.64^{+0.08}_{-0.06}$\\[5pt]
				\hline
			\end{tabular}
		
	}
	\caption{\scriptsize Parameters derived for the different parameters of the five models of the joint fit, described in Appendix \ref{appendix}.}
	\label{tab:res_joint}
	\end{center}
\end{table*}
		
		Previously published {\it XMM-Newton} X-ray to lensing mass ratios are obtained with a selection on lensing, and show values of $\left<M_{\rm {X}}/M_{\rm {L}}\right> \sim 0.91-0.99$ \citep{zhang2008} and $\sim 0.72-0.96$ \citep[\citet{simet2015}, using observations from ][]{piffaretti2011,hajian2013}.   Given that we measure the bias on the lensing mass given an X-ray selection, we cannot compare our measurements directly with those obtained by the measure of the bias in the X-ray mass given the lensing mass. However, the trend is similar and consistent with simulation. Our uncertainty on $\left<M_{\rm {L}}/M_{\rm {X}}\right>$ ($\sigma_{\rm  \left<M_{\rm {L}}/M_{\rm {X}}\right> } \sim15-20\%$) is also similar to those cited in these works ($\sigma_{\rm  \left<M_{\rm {X}}/M_{\rm {L}}\right> } \sim3-20\%$). 
		
		We remind the reader, however, that even if our results are consistent with previous work, the scaling relations, difference and ratios that we obtain between our lensing masses and those in  \citet{gozaliasl2014}  depend on the  \citet{gozaliasl2014} selection in $L_X$ (when stacking clusters to derive the \citet{leauthaud2010} lensing masses). Our selection based on the \citet{licitra2016a,licitra2016b} richness and differences in the shear calibration in our data and  \citet{leauthaud2010} contribute to this difference; interpreting them precisely implies understanding the degeneracies on each contribution.
		
%	There is a known tension between the constrains on the cosmological parameters derived using the number density of S-Z galaxy clusters detected by Planck, and those derived from the CMB temperature power spectrum. Planck cluster masses are derived with hydrostatic mass measurements applied to XMM-Newton X-ray observations. One possible explanation of this discrepancy could then be that these masses are biased low with respect to true masses. Our results are consistent with this explanation.  {\bf To reconcile the constrains on the cosmological parameters derived using the number density of S-Z galaxy clusters detected by Planck, and those derived from the CMB temperature power spectrum, the Planck S-Z masses, which are calibrated on X-ray masses, requires that X-ray mass estimation are $0.58\pm 0.04$ lower than true cluster masses \citep{planck2016}. We find lensing masses higher of $\sim 15\%$, and confirm the conclusion that this tension is most probably due to the fact that Planck masses are underestimated because of their calibration on X-ray masses \citep{postman2012, vonderlinden2014, hoekstra2015, pennalima2016, amodeo2017}}.
		
		It is also known that {\it XMM-Newton} and \textit{Chandra} have different instrument calibrations that lead to different temperature estimations, with  \textit{Chandra} X-ray temperatures being higher and leading to higher cluster mass estimation \citep{israel2014, vonderlinden2014, schellenberger2015}. Applying the correction from \citet{schellenberger2015}, to convert {\it XMM-Newton} masses to \textit{Chandra} masses, we find $\left<M_{\rm {L}}/M_{\rm {X}}\right>_{\rm Chandra}=0.99\pm0.17$, using the lensing masses from our {\it Final Model}.

\section{SUMMARY AND CONCLUSIONS}

	We measure weak lensing galaxy cluster masses for optically detected cluster candidates stacked by richness. We fit the weak lensing mass versus richness relation and compare our findings to X-ray detected mass proxies in the area.

	Our cluster sample was obtained with the RedGOLD \citep{licitra2016a} optical cluster finder algorithm. The algorithm is based on a revised red-sequence technique and searches for passive ETG overdensities. RedGOLD is optimized to detect massive clusters ( $M_{\rm 200}>10^{14}M_{\rm \odot}$) with both high completeness and purity. We use the RedGOLD cluster catalogs from \citet{licitra2016a, licitra2016b}  for the CFHT-LS W1 and NGVS surveys. The catalogs give the detection significance and an optical richness estimate that corresponds to a proxy for the cluster mass. 

	For our weak lensing analysis, we use a sample of 1323 published clusters, selected with a threshold in significance of $\sigma_{\rm det}\ge4$ and in richness $\lambda\ge10$ at redshift $0.2\le z\le0.5$, for which our published catalogs are $\sim 100\%$ complete and $\sim 80\%$ pure \citep{licitra2016a}. In order to compare our lensing masses to X-ray mass proxies, we considered both the {\it published} and {\it complete} Licitra et al.'s catalogs, as defined in Section \ref{redgold_cat}.
	
	Our photometric and photometric redshift catalogs were obtained with a modified version of the THELI pipeline \citep{erben2005, erben2009, erben2013, raichoor2014}, and weak lensing shear measurements with the shear measurement pipeline described in \citet{erben2013}, \citet{heymans2012}, and \citet{miller2013}.
	
	We calculate our cluster mean shear radial profiles by averaging the tangential shear in logarithmic radial bins in stacked cluster detections binned by their richness. We apply lens-source pairs weights that depend on the lensing efficiency and on the quality of background galaxy shape measurements. 

	We obtain the average cluster masses in each richness bin by fitting the measured shear profiles using three models: (1) a basic halo model ({\it Basic Model}), with an NFW surface density contrast and correction terms that take into account cluster miscentering, non-weak shear, and the second halo term; (2) a model that includes the intrinsic scatter in the mass--richness relation ({\it Added Scatter Model}); and (3) a model that includes the contribution of the BCG stellar mass ({\it Two Component Model}). In the {\it Basic} and in the {\it Two Component Models}, we apply an a posteriori correction to take into account the intrinsic scatter in the mass--richness relation. 
	
	We find that our {\it Final Model} is the {\it Two Component Model}, which, with the inclusion of the a posteriori correction for the intrinsic scatter in the mass--richness relation, is more complete in taking into account the systematics, and more reliable in the obtained results. 
	
	Our main results are:
	
	 \begin{itemize}
	 	
	 	\item We test different cluster profile models and fitting techniques. We find that the intrinsic scatter in the mass--richness relation and the BCG mass are not constrained by the data. While the miscentering correction is necessary to avoid a bias in the measured halo masses, the inclusion of the BCG mass does not affect the results. 
	 	
	 	\item Comparing weak lensing masses to RedGOLD optical richness, we calibrate our optical richness with the lensing masses, fitting the power law $\log{M_{\rm 200}}=\log{M_{\rm 0}}+\alpha\log{\lambda/\lambda_{\rm 0}}$. For our {\it Final Model}, we obtain $\log{M_{\rm 0}}=14.46\pm0.02$ and $\alpha=1.04\pm0.09$, with a pivot richness $\lambda_{0}=40$. 
		
			Even if our sample is one order of magnitude smaller than the SDSS and DES redMaPPer cluster samples used in  \citet{simet2016}, \citet{farahi2016} and \citet{melchior2016}, our results are consistent with theirs within $1-2 \sigma$. This confirms that our cluster selection is not biased toward a different cluster selection when compared to the SDSS and DES redMaPPer cluster samples, as we expect. 

	 	\item Using our mass--richness relation and X-ray masses from  \citet{gozaliasl2014}, we infer scaling relations between lensing masses and X-ray proxies. 
	 
	 		For the lensing mass versus X-ray luminosity relation $\log{\left(\frac{M_{\rm 200} E(z)}{M_{\rm 0}}\right)} =a+b \log{\left(\frac{L_{\rm X}}{L_{\rm 0} E(z)}\right)}$, we find $a=(0.10\pm0.03)$ and $b=(0.61\pm0.12)$, with $M_{\rm 0}=8 \times 10^{13} h^{-1}M_{\rm \odot}$ and $L_{\rm 0}=5.6\times 10^{42} h^{-2}erg/s$. 
	 
	 		For the lensing mass versus X-ray temperature relation $\log{\left(\frac{M_{\rm 200} E(z)}{M_{\rm 0}}\right)}=a+b \log{\left(\frac{T_{\rm X}}{T_{\rm 0}}\right)}$, we obtain $a=(0.23\pm0.03)$ and $b=(1.47\pm0.28)$, with $M_{\rm 0}=6 \times 10^{13} h^{-1}M_{\rm \odot}$ and $T_{\rm 0}=1.5 KeV$. 
	 
	 		Our results are consistent with those of \citet{kettula2015} and \citet{mantz2016}, within $<1\sigma$. Our normalization is consistent within $<2.5\sigma$, and our slope within $1\sigma$, of the results of \citet{leauthaud2010} (and therefore with \citet{gozaliasl2014}). They are also consistent with expected deviations from self-similarity \citep{bohringer2011}.
	 
		\item We find a scatter of $0.20$~dex, for all three relations, consistent with redMaPPer scatters, confirming the \citet{licitra2016a,licitra2016b} results that the RedGOLD optical richness is an efficient mass proxy. This is very promising because our mass range is lower than that probed by redMaPPer, and the scatter does not increase as expected to these lower mass ranges.

	\end{itemize}
	 In order to increase the accuracy of the weak lensing mass estimates, it will be important to increase the number density of background sources to achieve a higher S/N in the shear profile measurements in the future.
	 This will be possible with ground- and space-based large-scale surveys such as the LSST\footnote{https://www.lsst.org/}, \textit{Euclid}\footnote{http://euclid- ec.org}  and WFIRST\footnote{http://wfirst.gsfc.nasa.gov}. Also, the next generation radio surveys such as SKA\footnote{http://www.skatelescope.org} will allow us to extend weak lensing measurements to the radio band, giving access to even larger scales. Cluster samples will then be an order of magnitude bigger than the one used for this work, allowing us to constrain cluster masses and their scaling relations with even higher accuracy (e.g. \citet{sartoris2016}, \citet{ascaso2016}).\\

\acknowledgments
This work is based on observations obtained with MegaPrime/MegaCam, a joint project of CFHT and CEA/IRFU, at the Canada-France-Hawaii Telescope (CFHT) which is operated by the National Research Council (NRC) of Canada, the Institut National des Sciences de l'Univers of the Centre National de la Recherche Scientifique (CNRS) of France, and the University of Hawaii. This research used the facilities of the Canadian Astronomy Data Centre, operated by the National Research Council of Canada with the support of the Canadian Space Agency. CFHTLenS data processing was made possible thanks to significant computing support from the NSERC Research Tools and Instruments grant program.
R.L.,  S.M., and A.Ra. acknowledge the support of the French Agence Nationale
de la Recherche (ANR) under the reference ANR10-
BLANC-0506-01-Projet VIRAGE (PI: S.Mei). S.M. acknowledges financial support from the Institut Universitaire de France (IUF), of which she is senior member.
H.H. is supported by the DFG Emmy Noether grant Hi 1495/2-1. We thank the Observatory of Paris and the University of Paris D. Diderot for hosting T.E. under their visitor programs.

{\it Facilities:} \facility{CFHT}.

\begin{appendices}
\section{JOINT FIT TEST}\label{appendix}
In order to check that individually fitting the profile of each richness bin does not introduce a bias in the determination of the mass--richness relation parameters, we tested a \textit{joint fit}  \citep[e.g.][]{viola2015, simet2016}. This method consists of the simultaneous fitting of the profiles associated whit all richness bins. In this case, the fitting parameters will be directly the normalization and slope of the mass--richness relation, and the likelihood of the model will be the sum of the likelihoods of all shear profiles.
	
Also for the \textit{joint fit}, we changed the free parameters to understand how each free parameter could change the results. We tested different models, each with different free parameters:
	
\begin{description}
		\item[Model~1] has four parameters: $\log M_{0}$, $\alpha$, $p_{\rm cc}$, and $\sigma_{\rm off}$, which are the normalization and slope of the mass--richness relation, and the miscentering parameters. The BCG mass is fixed at $M_{\rm BCG}^{*}$.
		
		\item[Model~2] has five parameters: $\log M_{0}$, $\alpha$, $p_{\rm cc}$, $\sigma_{\rm off}$, and $\sigma_{M|\lambda}$, which are the parameters of Model~1 with the addition of the intrinsic scatter of the mass--richness relation. The BCG mass is fixed at $M_{\rm BCG}^{*}$.
		
		\item[Model~3] has five parameters: $\log M_{0}$, $\alpha$, $p_{\rm cc}$, $\sigma_{\rm off}$, and $aMbcg$, which are the parameters of Model~1 with the addition of a constant that multiplies $M_{\rm BCG}^{*}$, so that $M_{\rm BCG}=aMbcg\times M_{\rm BCG}^{*}$.
		
		\item[Model~4] has five parameters: $\log M_{0}$, $\alpha$, $p_{\rm cc}$, $\sigma_{\rm off}$, and $aCM$, which are the parameters of Model~1 with the addition of the amplitude of the mass-concentration relation used \citep[i.e.][]{dutton2014}. The BCG mass is fixed at $M_{\rm BCG}^{*}$.
				
		\item[Model~5] has seven parameters: $\log M_{0}$, $\alpha$, $p_{\rm cc}$, $\sigma_{\rm off}$, $\sigma_{M|\lambda}$, $aMbcg$, and $aCM$.
	
\end{description}  

In Table \ref{tab:priors_joint} we find the priors on the parameters. In Table \ref{tab:res_joint}, we find the results of the MCMC for the different models. 
	
We found that the results from the different models are consistent with each other within $<1.5\sigma$; except one, the $\log M_{0}$ obtained with Model~4, which is only consistent with those from Models~1 and 3 within $2.2\sigma$ . The normalization and slope of the mass--richness relation are well-constrained in all models. Here, $aMbcg$ is not constrained, and the inclusion of this parameter in the fit does not affect the other parameters. This result is consistent with what was found in Section \ref{results}, from the comparison of the {\it Basic} and {\it Two Component Models}. The amplitude of mass--concentration relation is constrained, but it is slightly degenerate with the miscentering parameters that are less well-constrained in the models that include $aCM$. Moreover, for these models, the normalization and slope of the mass--richness relation have lower values compared to the models without $aCM$. Here, $\sigma_{M|\lambda}$ is constrained but it has a lower value than expected from \citet{licitra2016a, licitra2016b}.
	
When comparing the results from the \textit{joint fit} to the results from our {\it Final Model}, we find consistent results ($<1-2\sigma$), confirming that the two approaches are consistent and equivalent.

\end{appendices}

\end{document}